\documentclass[11pt]{article}
\usepackage[T1]{fontenc}
\usepackage{lmodern}
\usepackage[margin=1in]{geometry}
\usepackage{amsmath,amssymb,amsthm}
\usepackage{booktabs,array,enumitem}
\usepackage{caption}
\usepackage{microtype}
\usepackage{xurl}
\usepackage{cite}
\usepackage[hidelinks]{hyperref}
\newtheorem{theorem}{Theorem}[section]
\newtheorem{lemma}[theorem]{Lemma}
\newtheorem{proposition}[theorem]{Proposition}
\newtheorem{corollary}[theorem]{Corollary}
\theoremstyle{definition}
\newtheorem{definition}[theorem]{Definition}
\newtheorem{schurdefinition}[theorem]{Definition}
\theoremstyle{remark}
\newtheorem{remark}[theorem]{Remark}
\DeclareMathOperator{\rank}{rank}

\begin{document}
\title{Maximal Hull Dimensions of a Class of Twisted Roth--Lempel Codes under Monomial Equivalence
\thanks{This work was supported by the National Natural Science Foundation of China (Grant No.~2441105) and the Interdisciplinary Project of Capital Normal University (Grant No.~2026JCYB02).}}
\author{Huiyue~Lei\thanks{Huiyue Lei is with the School of Mathematical Sciences, Capital Normal University, Beijing 100048, China. Email: 2250501037@cnu.edu.cn.},
Haojie~Gu\thanks{Haojie Gu is with the School of Mathematical Sciences, Capital Normal University, Beijing 100048, China. Email: 2200502051@cnu.edu.cn.},
and Jun~Zhang\thanks{Jun Zhang is with the School of Mathematical Sciences, Capital Normal University, Beijing 100048, China. Email: junz@cnu.edu.cn. Corresponding author.}}
\date{}
\maketitle

\begin{abstract}
We study how coordinate rescaling affects the hull dimension
of twisted Roth--Lempel codes with hook position $\ell=k-2$
and nonzero twist.
We consider Euclidean hulls over $\mathbb F_q$ for even $q$
and Hermitian hulls after scalar extension to $\mathbb F_{q^2}$
for arbitrary prime powers $q$.
For each fixed pair of multipliers on the two appended
coordinates, the attainable hull dimensions form an interval
starting at zero.
We determine its upper endpoint in several parameter ranges
by minimizing the rank of a Gram matrix.
As an application, we use the maximal hull dimension under
monomial equivalence to distinguish a family of non-GRS
MDS codes from twice-extended TGRS codes, even for pairs
with identical parameters and full Schur squares.
\end{abstract}

\begin{flushleft}
\end{flushleft}

\section{Introduction}\label{sec:intro}

Let $q$ be a prime power, and write
$\mathbb F_q^*=\mathbb F_q\setminus\{0\}$. 
The Schur square of a linear code $C\subseteq\mathbb F_q^N$ is the span of the componentwise
products of pairs of its codewords. Its dimension is invariant under
coordinate permutations and nonzero coordinate scalings. Suppose two
maximum distance separable (MDS) codes have the same parameters and
their Schur squares both fill the ambient space. What other invariant
can distinguish their monomial equivalence classes? One such invariant is the maximal Euclidean hull dimension, the largest
dimension that the intersection of a code with its Euclidean dual can
attain under coordinate rescaling.

The Euclidean dual of a linear code $C\subseteq\mathbb F_q^N$
of length $N$ is
\[
C^\perp=
\left\{\boldsymbol x\in\mathbb F_q^N:
\sum_{i=1}^N x_i c_i=0
\text{ for every }\boldsymbol c\in C\right\}.
\]
For a linear code $C\subseteq\mathbb F_{q^2}^N$, its Hermitian dual is
\[
C^{\perp_H}=
\left\{\boldsymbol x\in\mathbb F_{q^2}^N:
\sum_{i=1}^N x_i c_i^q=0
\text{ for every }\boldsymbol c\in C\right\}.
\]
The corresponding hulls are~\cite{assmus1990planes}
\[
\operatorname{Hull}_E(C)=C\cap C^\perp,
\qquad
\operatorname{Hull}_H(C)=C\cap C^{\perp_H}.
\]
A code is called Euclidean or Hermitian linear complementary dual
(LCD) if its corresponding hull is $\{0\}$~\cite{massey1992lcd}.
A hull equal to $C$ characterizes self-orthogonality under the
corresponding inner product.
Hull dimensions play a role in algorithms for code
equivalence~\cite{sendrier2000support} and in constructions of
entanglement-assisted quantum error-correcting
codes~\cite{luo2019hulls,fang2020hulls,gao2021goppa}.

Coordinate permutations and nonzero coordinate scalings define
monomial equivalence and preserve the length, dimension, and
minimum distance of a code.
Coordinate scalings can nevertheless change its hull dimension.
Carlet et al.~\cite{carlet2018equiv} proved that every linear code
over $\mathbb F_q$ is monomially equivalent to a Euclidean LCD
code when $q>3$, and every linear code over $\mathbb F_{q^2}$ is
monomially equivalent to a Hermitian LCD code when $q>2$.
Under these assumptions, every dimension between zero and the
hull dimension of a given representative is attainable by a
monomially equivalent code~\cite[Corollary~3.9]{anderson2024relative}.
Chen~\cite{chen2023variation} introduced the maximal Euclidean and
Hermitian hull dimensions over a monomial equivalence class and
studied them as invariants of that class.

Generalized Reed--Solomon (GRS) codes are a basic class of maximum
distance separable (MDS) codes.
A code is called non-GRS here if it is not monomially equivalent
to a GRS code.
Roth--Lempel codes~\cite{roth1989construction} and twisted
Reed--Solomon codes~\cite{beelen2022twisted} provide constructions
of non-GRS MDS codes.
Broader classes of twisted generalized Reed--Solomon (TGRS) codes
have been studied in~\cite{gu2024ltwists,hu2025LP,liu2026generalhook},
and their parity-check matrices were investigated
in~\cite{cheng2024parity}.

Hull properties of these families have also received attention.
LCD constructions and codes with small hull dimensions appear
in~\cite{wu2021lcdnonrs,wu2021onedimhull,singh2024multitwisted,liang2026fourlcd},
while self-orthogonality and Hermitian hulls are considered
in~\cite{liu2026generalhook,jia2026hermitian,sok2022hulls}.
For generalized Roth--Lempel codes,
Liang et al.~\cite{liang2026grlhull} obtained LCD constructions
and bounds on attainable hull dimensions for certain classes.
Liu et al.~\cite{liu2026grlselforth} established criteria for
Hermitian self-orthogonality, while
Wu et al.~\cite{wu2026galoishull} constructed codes with prescribed
Galois hull dimensions under conditions on the
appended-coordinate matrix.

The twisted Roth--Lempel (TRL) codes introduced
in~\cite{lei2026trl} combine evaluation of a single-twisted
polynomial space with two appended coefficient coordinates.
That work studies minimum distance, MDS and near-MDS criteria,
and non-GRS properties.
The twice-extended TGRS codes of Cheng and
Zhu~\cite{cheng2026twice} are monomially equivalent to the
$\ell=k-1$ specialization of the TRL family, as explained in
Subsection~\ref{sec:hook-comparison}.
Their work includes parity-check matrices, MDS and near-MDS
conditions, non-GRS properties, and several LCD constructions.
Building on these structural and hull results, we determine exact
hull maxima for the specialization $\ell=k-2$, including
constraints on the multipliers of the two appended coordinates.

We use the exact hull maxima to compare MDS codes from the
$\ell=k-2$ TRL and twice-extended TGRS families when their
parameters and Schur-square dimensions agree.
The construction in Section~\ref{sec:codes} shows that maximal
hull dimension can distinguish such codes.

\subsection{Motivation and problem statement}

Chen~\cite{chen2023variation} introduced the maximal
Euclidean and Hermitian hull dimensions as invariants
under monomial equivalence.
Lao et al.~\cite{lao2026maximum} studied bounds on the
maximum Hermitian hull dimension among all MDS codes
with prescribed length and dimension.
These results provide a general framework and bounds
for maximal hull dimensions, but do not directly determine
the maximum within the monomial equivalence class
of a fixed TRL code.
We therefore investigate the maximal hull dimension
obtainable by nonzero coordinate scalings of a fixed
$[n+2,k]_q$ TRL code with $\ell=k-2$ and a
nonzero twist parameter, keeping its evaluation
points and defining parameters fixed
(Definition~\ref{def:TRL-code}).

We consider Euclidean hulls over $\mathbb F_q$
when $q$ is even, and Hermitian hulls after scalar
extension from $\mathbb F_q$ to $\mathbb F_{q^2}$
for arbitrary prime powers $q$.
Let $a,b$ be the nonzero multipliers of the two
appended coordinates, and define
\[
\tau=
\begin{cases}
(a/b)^2, & \text{in the Euclidean case},\\
N(a/b), & \text{in the Hermitian case},
\end{cases}
\]
where $N(z)=z^{q+1}$.

By Proposition~\ref{prop:restricted-scaling-maximal-hull},
the largest hull dimension obtainable by varying the
evaluation-coordinate multipliers depends on the two prescribed
multipliers only through $\tau$.
We denote this maximum by $h^{(k-2)}(\tau)$.
The Euclidean and Hermitian maxima agree when $q$ is even.
Allowing all coordinate multipliers to vary gives
\[
h_{\max}^{(k-2)}
=\max_{\tau\in\mathbb F_q^*}h^{(k-2)}(\tau).
\]
Since coordinate permutations preserve hull dimension,
$h_{\max}^{(k-2)}$ is the maximum over the full monomial
equivalence class.
The central question is how the hull dimension of a fixed
TRL code varies within its monomial equivalence class.
More specifically, for prescribed $\tau$, what is the
largest hull dimension attainable by rescaling the
evaluation coordinates, and which hull dimensions remain
attainable when the multipliers of the two appended coordinates
are fixed?

\subsection{Main results and application}

Our main results and their application are as follows.
\begin{itemize}[leftmargin=*,itemsep=0.4\baselineskip,
  topsep=0.4\baselineskip,parsep=0pt]
\item
\emph{Maximal and attainable hull dimensions.}
With $r=n-k$, we determine $h^{(k-2)}(\tau)$ for $k\ge4$ and
$r\ge2$ in the parameter ranges and under the field-size conditions listed in
Table~\ref{tab:scope}.
In particular, for $r=k-1$ and $k\ge5$, we give necessary and
sufficient conditions for $h^{(k-2)}(\tau)$ to equal $k$,
$k-1$, or $k-2$, without an additional field-size assumption
(Theorem~\ref{thm:kminus1-exact-maxima}).
For $r=k-1$ and all $k\ge4$, we also characterize when the maximum
over $\tau$ equals $k$ or $k-1$
(Corollary~\ref{cor:max-hull-kminus1}).

For every $k\ge4$ and $n>k$,
Corollary~\ref{cor:fixed-tau-attainable-hulls} shows that every
integer between zero and $h^{(k-2)}(\tau)$ is attainable while
the multipliers of the two appended coordinates remain fixed.
Combined with the formulas for the maximum, this determines all
attainable hull dimensions in the cases listed in
Table~\ref{tab:scope}.

\item
\emph{Application to MDS constructions and monomial inequivalence.}
For codes of length $2k$ with $r=k-2\ge3$,
Theorem~\ref{thm:first-constrained-lower} gives a criterion for
attaining hull dimension $k-1$.
We use this criterion to construct
explicit non-GRS $[2k,k,k+1]_q$ MDS codes with
$h_{\max}^{(k-2)}=k-1$.
The construction in
Theorem~\ref{thm:explicit-nongrs-mds-family} uses
$q=q_0^m$, where $q_0=2^s\ge16$, $m\ge3$, and gives explicit coordinate multipliers attaining
the maximal hull dimension $k-1$.
A separate calculation shows that, for the same $k$ and $q$,
every $[2k,k]_q$ code in the twice-extended TGRS family of
Cheng and Zhu has maximal hull dimension $k-2$
(Proposition~\ref{prop:hook-kminus1-realization-threshold}).
Hence none of the constructed TRL codes is monomially equivalent
to a twice-extended TGRS code with these parameters $[2k,k,k+1]_q$.

For each constructed code, changing $\ell$ from $k-2$ to $k-1$
while keeping the evaluation points and the remaining defining
parameters unchanged gives an MDS code with the same
$[2k,k,k+1]_q$ parameters
(see Subsection~\ref{sec:hook-comparison}).
Both codes have Schur-square dimension $2k$
(Proposition~\ref{prop:balanced-schur-squares}).
Thus maximal hull dimension distinguishes these codes even though
their parameters and Schur-square dimensions coincide.
The maximal Hermitian hull dimensions of their scalar extensions
are also $k-1$ and $k-2$, respectively, so the codes remain
monomially inequivalent over $\mathbb F_{q^2}$.
\end{itemize}

\begin{center}
\begin{minipage}{\linewidth}
\captionsetup{hypcap=false,font=small,labelfont=bf,
  justification=raggedright,singlelinecheck=false}
\captionof{table}{Maximal hull dimensions $h^{(k-2)}(\tau)$
for prescribed $\tau$, with $k\ge4$ and $r\ge2$.}
\label{tab:scope}
\small
\setlength{\tabcolsep}{5pt}
\renewcommand{\arraystretch}{1.3}
\begin{tabular*}{\linewidth}{@{\extracolsep{\fill}}
  >{\raggedright\arraybackslash}p{0.23\linewidth}
  >{\centering\arraybackslash}p{0.23\linewidth}
  >{\centering\arraybackslash}p{0.26\linewidth}
  >{\raggedleft\arraybackslash}p{0.20\linewidth}@{}}
\multicolumn{4}{@{}l@{}}{\textbf{I. $2\le r\le k-2$: using parity-check matrices}}\\[3pt]
\toprule
\textbf{Range} & \textbf{Sufficient condition on $q$} &
$\boldsymbol{h^{(k-2)}(\tau)}$ & \textbf{Result}\\
\midrule
$2\le r\le k-4$ & $q>n$ &
  $3$ if $r=2$\newline $r,\ r+1$ if $r\ge3$
  & Thm.~\ref{thm:free-maximum-hulls}\\
$r=k-3\ge3$ & $q>n$ & $r,\ r+1$
  & Thm.~\ref{thm:critical-lower-exact-rank}\\
$(k,r)=(5,2)$ & $q>n$ & $2,\ 3$
  & Thm.~\ref{thm:k5-r2-exact}\\
$r=k-2\ge3$ & $q>2n$ & $r,\ r+1$
  & Thm.~\ref{thm:first-constrained-lower}\\
\bottomrule
\addlinespace[10pt]
\multicolumn{4}{@{}l@{}}{\textbf{II. $r\ge k-1$: using generator matrices}}\\[3pt]
\toprule
\textbf{Range} & \textbf{Sufficient condition on $q$} &
$\boldsymbol{h^{(k-2)}(\tau)}$ & \textbf{Result}\\
\midrule
$r=k-1$, $k\ge5$ & $q\ge n$ & $k,\ k-1,\ k-2$
  & Thm.~\ref{thm:kminus1-exact-maxima}\\
$r=k$, $k\ge4$ & $q>2n$ & $k-1,\ k$
  & Thm.~\ref{thm:generator-maxima}\textup{(i)}\\
$r=k+1$, $k\ge4$ & $q\ge n$ & $k-1,\ k$
  & Thm.~\ref{thm:generator-maxima}\textup{(ii)}\\
$r\ge k+2$, $k\ge4$ & $q>n$ & $k$
  & Thm.~\ref{thm:generator-maxima}\textup{(iii)}\\
\bottomrule
\end{tabular*}

\par\smallskip

\noindent\emph{Note.}
The condition $q\ge n$ allows the choice of $n$ distinct
evaluation points in $\mathbb F_q$.
In the rows $r=k-2\ge3$ and $r=k\ge4$, the criteria for the
larger value hold for all $q\ge n$.
The bound $q>2n$ guarantees the smaller value when the
corresponding criterion fails
(Theorems~\ref{thm:first-constrained-lower}
and~\ref{thm:generator-maxima}).
The fixed-$\tau$ classification excludes
$(k,r)=(4,2)$ and $(4,3)$, although the
maximum over $\tau$ is determined for
$(4,3)$ by Corollary~\ref{cor:max-hull-kminus1}.

\end{minipage}
\end{center}

Section~\ref{sec:prelim} presents the TRL codes, scaling
normalization, and attainability with fixed appended-coordinate
multipliers.
Section~\ref{sec:hankel} develops the moment representation and
perturbation tools.
Sections~\ref{sec:dual-maxima} and~\ref{sec:generator} determine
the maximal hull dimensions using parity-check and generator
matrices, respectively.
Section~\ref{sec:codes} applies the criterion at length $2k$ to
MDS constructions and proves monomial inequivalence to the
twice-extended TGRS comparison family.
Section~\ref{sec:conclusion} concludes the paper.
Appendix~\ref{app:k5-r2} treats $(k,r)=(5,2)$.
\section{Preliminaries}\label{sec:prelim}
\subsection{TRL codes and matrix representations}
\label{sec:gram-conventions}

We recall the following definition of twisted
Roth--Lempel codes~\cite{lei2026trl}.

\begin{definition}\label{def:TRL-code}
Let $3\le k<n\le q$, $0\le\ell\le k-1$, and $\eta,\delta\in\mathbb F_q$. Define
\[
S_{k,\ell,\eta}
=
\left\{
 f(x)=\sum_{i=0}^{k-1}f_ix^i+\eta f_\ell x^k:
 f_0,\ldots,f_{k-1}\in\mathbb F_q
\right\}.
\]
For pairwise distinct $\alpha_1,\ldots,\alpha_n\in\mathbb F_q$, write
$\boldsymbol\alpha=(\alpha_1,\ldots,\alpha_n)$ and
$\mathcal A=\{\alpha_1,\ldots,\alpha_n\}$. Define
\[
\operatorname{TRL}_{k,n+2}(\boldsymbol\alpha,\eta,\delta,\ell)
=
\left\{
(f(\alpha_1),\ldots,f(\alpha_n),f_{k-1},f_{k-2}+\delta f_{k-1}):
 f\in S_{k,\ell,\eta}
\right\}.
\]

\end{definition}

Since $\deg f\le k<n$, evaluation at the $n$ distinct points is
injective on $S_{k,\ell,\eta}$. Thus
\[
\mathcal C^{(\ell)}
=\operatorname{TRL}_{k,n+2}(\boldsymbol\alpha,\eta,\delta,\ell)
\]
is an $[n+2,k]_q$ linear code. Its scalar extension to
$\mathbb F_{q^2}$ is denoted by
\[
\mathcal C_{q^2}^{(\ell)}
=\mathbb F_{q^2}\otimes_{\mathbb F_q}\mathcal C^{(\ell)}.
\]
For $\boldsymbol v=(v_1,\ldots,v_n)\in(\mathbb F_q^*)^n$, put
\[
\mathcal C_{\boldsymbol v}^{(\ell)}
=\mathcal C^{(\ell)}\operatorname{diag}(v_1,\ldots,v_n,1,1).
\]

For $\ell=k-2$, a generator matrix of
$\mathcal C_{\boldsymbol v}^{(k-2)}$ is
\begin{equation}
\label{eq:generator-lower}
G_{k-2}=
\begin{pmatrix}
 \bigl(v_i\alpha_i^j\bigr)_{\substack{0\le j\le k-3\\1\le i\le n}}
 &\mathbf0_{(k-2)\times1}&\mathbf0_{(k-2)\times1}\\[3pt]
 \bigl(v_i(\alpha_i^{k-2}+\eta\alpha_i^k)\bigr)_{1\le i\le n}&0&1\\[2pt]
 \bigl(v_i\alpha_i^{k-1}\bigr)_{1\le i\le n}&1&\delta
\end{pmatrix}.
\end{equation}
For $\ell=k-1$, a generator matrix of
$\mathcal C_{\boldsymbol v}^{(k-1)}$ is
\begin{equation}
\label{eq:generator-upper}
G_{k-1}=
\begin{pmatrix}
 \bigl(v_i\alpha_i^j\bigr)_{\substack{0\le j\le k-3\\1\le i\le n}}
 &\mathbf0_{(k-2)\times1}&\mathbf0_{(k-2)\times1}\\[3pt]
 \bigl(v_i\alpha_i^{k-2}\bigr)_{1\le i\le n}&0&1\\[2pt]
 \bigl(v_i(\alpha_i^{k-1}+\eta\alpha_i^k)\bigr)_{1\le i\le n}&1&\delta
\end{pmatrix}.
\end{equation}
In each matrix, the indexed tuples in the last two rows are row vectors.

For a matrix $A$ over $\mathbb F_{q^2}$, let
$A^\dagger=\overline A^{\mathsf T}$, where $\overline A$ is obtained
by raising each entry of $A$ to the $q$th power.
The Gram matrices for the Euclidean and Hermitian products are
$AA^{\mathsf T}$ and $AA^\dagger$, respectively.

Let $G$ and $H$ be generator matrices of an
$[n+2,k]_q$ code $C$ and its Euclidean dual,
respectively. The hull dimension satisfies~\cite{fang2020hulls,chen2023variation}
\[
\dim\operatorname{Hull}_E(C)
=k-\rank(GG^{\mathsf T})
=n+2-k-\rank(HH^{\mathsf T}).
\]
For an $[n+2,k]_{q^2}$ code $C$, let $G$ and
$H_H$ be generator matrices of $C$ and
$C^{\perp_H}$, respectively. Then
\[
\dim\operatorname{Hull}_H(C)
=k-\rank(GG^\dagger)
=n+2-k-\rank(H_HH_H^\dagger).
\]
In particular, $C$ is LCD if and only if
its corresponding generator Gram matrix is
nonsingular.

To describe parity-check matrices of
$\mathcal C_{\boldsymbol v}^{(\ell)}$, define
\[
Q(x)=\prod_{i=1}^n(x-\alpha_i)
=x^n+\sigma_1x^{n-1}+\sigma_2x^{n-2}+\cdots+\sigma_n.
\]
Thus $\sigma_j=(-1)^j e_j(\alpha_1,\ldots,\alpha_n)$, where $e_j$
is the $j$th elementary symmetric polynomial. Set
\[
u_i=Q'(\alpha_i)^{-1}
=\left(\prod_{\substack{1\le j\le n\\j\ne i}}(\alpha_i-\alpha_j)\right)^{-1},
\qquad 1\le i\le n.
\]
For $s\ge0$, set
\[
w_s=\sum_{i=1}^n u_i\alpha_i^s.
\]

The following standard Lagrange identities are recorded
in~\cite{liang2026fourlcd}.
\begin{lemma}\label{lem:lagrange-moments}
The sums $w_s$ satisfy
\[
\begin{gathered}
w_s=0\quad\text{for }0\le s\le n-2,\qquad
w_{n-1}=1,\qquad
w_n=-\sigma_1,\\
w_{n+1}=\sigma_1^2-\sigma_2,\qquad
w_{n+2}=-\sigma_1^3+2\sigma_1\sigma_2-\sigma_3.
\end{gathered}
\]
Consequently,
\[
\sum_{i=1}^n u_i f(\alpha_i)=0
\]
for every $f\in\mathbb F_q[x]$ with $\deg f\le n-2$.
\end{lemma}

The Lagrange identities in Lemma~\ref{lem:lagrange-moments}
yield explicit parity-check matrices for the
coordinate-scaled TRL codes, as follows.

\begin{lemma}\label{lem:normalized-parity-check}
Let $r=n-k$, $\eta\in\mathbb F_q^*$, and
$\boldsymbol v\in(\mathbb F_q^*)^n$.
Put $\Lambda_2=\sigma_1^2-\sigma_2$.
For $\ell\in\{k-2,k-1\}$, the code
$\mathcal C_{\boldsymbol v}^{(\ell)}$ has a parity-check matrix
\[
\widehat H_{\boldsymbol v}^{(\ell)}=[P_{\boldsymbol v}\mid F_\ell],
\qquad
P_{\boldsymbol v}
=
\left(\frac{u_i}{v_i}\alpha_i^j\right)_{
\substack{0\le j\le r+1\\1\le i\le n}},
\]
where $F_{k-2},F_{k-1}\in\mathbb F_q^{(r+2)\times2}$ are
\begin{equation}
F_{k-2}=
\begin{pmatrix}
\mathbf 0_{(r-1)\times1}&\mathbf 0_{(r-1)\times1}\\[2pt]
\eta\delta&-\eta\\
-1-\eta\delta\sigma_1&\eta\sigma_1\\
\delta+\sigma_1+\eta\delta\Lambda_2&-1-\eta\Lambda_2
\end{pmatrix}
\end{equation}
and
\begin{equation}
F_{k-1}=
\begin{pmatrix}
\mathbf 0_{(r-1)\times1}&\mathbf 0_{(r-1)\times1}\\[2pt]
-\eta&0\\
\eta\sigma_1-1&0\\
\delta+\sigma_1-\eta\Lambda_2&-1
\end{pmatrix}.
\end{equation}
When $r=1$, the zero blocks are absent.
\end{lemma}

For later use, we recall the Schur product and
the Schur square of a linear code
from~\cite[Definition~2.6]{cascudo2020squares}.

\begin{schurdefinition}
For vectors
$\boldsymbol x=(x_1,\ldots,x_N)$ and
$\boldsymbol y=(y_1,\ldots,y_N)$ in $\mathbb F_q^N$,
their \emph{Schur product} is defined by
\[
\boldsymbol x\star\boldsymbol y
=(x_i y_i)_{i=1}^N.
\]
For a linear code $C\subseteq\mathbb F_q^N$,
its \emph{Schur square} is
\[
C^{\star2}
=\operatorname{span}_{\mathbb F_q}
\{\boldsymbol x\star\boldsymbol y:
\boldsymbol x,\boldsymbol y\in C\}.
\]
\end{schurdefinition}

\subsection{Coordinate scalings and attainable hull dimensions}
\label{sec:scaling-normalization}
We use normalized Gram matrices to study maximal hull dimensions under
nonzero coordinate scalings and determine which hull dimensions are
attainable when the multipliers of the two appended coordinates are fixed.

Using the parity-check matrices in
Lemma~\ref{lem:normalized-parity-check},
we obtain the following result on the Gram matrices
of the dual codes under coordinate scalings.

\begin{proposition}\label{prop:restricted-scaling-maximal-hull}
Let $\eta\in\mathbb F_q^*$ and $\ell\in\{k-2,k-1\}$.
\begin{enumerate}[label=\textup{(\roman*)}]
\item If $q$ is even and $a,b\in\mathbb F_q^*$, then
\[
\max_{\boldsymbol v\in(\mathbb F_q^*)^n}
\dim\operatorname{Hull}_E
\bigl(\mathcal C^{(\ell)}\operatorname{diag}(v_1,\ldots,v_n,a,b)\bigr)
\]
depends on $a,b$ only through $\tau=(a/b)^2$.
\item If $a,b\in\mathbb F_{q^2}^*$, then
\[
\max_{\boldsymbol v\in(\mathbb F_{q^2}^*)^n}
\dim\operatorname{Hull}_H
\bigl(\mathcal C_{q^2}^{(\ell)}\operatorname{diag}(v_1,\ldots,v_n,a,b)\bigr)
\]
depends on $a,b$ only through $\tau=N(a/b)$, where $N(z)=z^{q+1}$.
\end{enumerate}
In either setting, after multiplication by a nonzero scalar, the dual
Gram matrices range over
\[
\left(\sum_{i=1}^n\rho_i\alpha_i^{s+t}\right)_{0\le s,t\le r+1}
+B^{(\ell)}(\tau),
\qquad \boldsymbol\rho\in(\mathbb F_q^*)^n,
\]
where
\[
B^{(\ell)}(\tau)=F_\ell\operatorname{diag}(1,\tau)F_\ell^{\mathsf T}.
\]
Every $\tau\in\mathbb F_q^*$ can be realized
by a suitable choice of $a,b$ in either setting.
\end{proposition}

\begin{proof}
In the Euclidean case, a parity-check matrix of
$\mathcal C^{(\ell)}
\operatorname{diag}(v_1,\ldots,v_n,a,b)$ is
\[
H_E=
\left[
P_{\boldsymbol v}\mid
-F_\ell\operatorname{diag}(a^{-1},b^{-1})
\right],
\]
where
\[
P_{\boldsymbol v}
=
\left(\frac{u_i}{v_i}\alpha_i^j\right)_{
\substack{0\le j\le r+1\\1\le i\le n}}.
\]
Its Gram matrix is
\[
\begin{aligned}
H_EH_E^{\mathsf T}
&=P_{\boldsymbol v}P_{\boldsymbol v}^{\mathsf T}
+F_\ell\operatorname{diag}(a^{-2},b^{-2})
F_\ell^{\mathsf T}\\
&=
\left(
\sum_{i=1}^n
\frac{u_i^2}{v_i^2}\alpha_i^{s+t}
\right)_{0\le s,t\le r+1}
+F_\ell\operatorname{diag}(a^{-2},b^{-2})
F_\ell^{\mathsf T}.
\end{aligned}
\]
Multiplying by $a^2$ gives
\[
\begin{aligned}
a^2H_EH_E^{\mathsf T}
&=
\left(
\sum_{i=1}^n
\frac{a^2u_i^2}{v_i^2}\alpha_i^{s+t}
\right)_{0\le s,t\le r+1}
+F_\ell\operatorname{diag}
\left(1,\frac{a^2}{b^2}\right)
F_\ell^{\mathsf T}.
\end{aligned}
\]
Thus the matrix has the stated form with
\[
\rho_i=\frac{a^2u_i^2}{v_i^2},
\qquad
\tau=(a/b)^2.
\]
For fixed $a$, the map
$v_i\mapsto a^2u_i^2/v_i^2$ is a bijection
of $\mathbb F_q^*$ when $q$ is even.
Consequently, the weights $\rho_1,\ldots,\rho_n$
can be prescribed arbitrarily in $(\mathbb F_q^*)^n$. Hence the minimum rank, and
therefore the maximum hull dimension, depends on
$a,b$ only through $\tau$.

In the Hermitian case, a generator matrix of the
Hermitian dual of
$\mathcal C_{q^2}^{(\ell)}
\operatorname{diag}(v_1,\ldots,v_n,a,b)$ is
\[
G_H=
\widehat H_{\boldsymbol 1}^{(\ell)}
\operatorname{diag}
(v_1^{-q},\ldots,v_n^{-q},a^{-q},b^{-q}).
\]
Since the entries of
$\widehat H_{\boldsymbol 1}^{(\ell)}$
belong to $\mathbb F_q$, its Hermitian Gram matrix is
\[
\begin{aligned}
G_HG_H^\dagger
&=
\left(
\sum_{i=1}^n
\frac{u_i^2}{N(v_i)}\alpha_i^{s+t}
\right)_{0\le s,t\le r+1}
+F_\ell\operatorname{diag}
\bigl(N(a)^{-1},N(b)^{-1}\bigr)
F_\ell^{\mathsf T}.
\end{aligned}
\]
Multiplying by $N(a)$ gives
\[
\begin{aligned}
N(a)G_HG_H^\dagger
&=
\left(
\sum_{i=1}^n
\frac{N(a)u_i^2}{N(v_i)}
\alpha_i^{s+t}
\right)_{0\le s,t\le r+1}
+F_\ell\operatorname{diag}
\left(1,\frac{N(a)}{N(b)}\right)
F_\ell^{\mathsf T}.
\end{aligned}
\]
Thus the same matrix form is obtained with
\[
\rho_i=\frac{N(a)u_i^2}{N(v_i)},
\qquad
\tau=\frac{N(a)}{N(b)}=N(a/b).
\]
The norm map
$N:\mathbb F_{q^2}^*\to\mathbb F_q^*$
is surjective, so the $\rho_i$ also range over $\mathbb F_q^*$.

Finally, the surjectivity of the square map in even
characteristic and of the norm map shows that every
$\tau\in\mathbb F_q^*$ can be realized in the
respective settings.
\end{proof}

For $\ell\in\{k-2,k-1\}$ and $\tau\in\mathbb F_q^*$,
define $h^{(\ell)}(\tau)$ as follows. In the Euclidean setting,
when $q$ is even, for any $a,b\in\mathbb F_q^*$ satisfying
$(a/b)^2=\tau$, let
\[
h^{(\ell)}(\tau)
=
\max_{\boldsymbol v\in(\mathbb F_q^*)^n}
\dim\operatorname{Hull}_E
\left(
\mathcal C^{(\ell)}
\operatorname{diag}(v_1,\ldots,v_n,a,b)
\right).
\]
In the Hermitian setting, for any $a,b\in\mathbb F_{q^2}^*$
satisfying $N(a/b)=\tau$, let
\[
h^{(\ell)}(\tau)
=
\max_{\boldsymbol v\in(\mathbb F_{q^2}^*)^n}
\dim\operatorname{Hull}_H
\left(
\mathcal C_{q^2}^{(\ell)}
\operatorname{diag}(v_1,\ldots,v_n,a,b)
\right).
\]
By Proposition~\ref{prop:restricted-scaling-maximal-hull},
these maxima depend on $a,b$ only through $\tau$.
The normalized Gram matrices have entries in $\mathbb F_q$
and the same rank over $\mathbb F_q$ and $\mathbb F_{q^2}$,
so the Euclidean and Hermitian maxima agree when $q$ is even.
Define
\begin{equation}\label{eq:monomial-hull-maximum}
h_{\max}^{(\ell)}
=\max_{\tau\in\mathbb F_q^*}h^{(\ell)}(\tau).
\end{equation}
Since coordinate permutations preserve hull dimension,
$h_{\max}^{(\ell)}$ is the maximal Euclidean hull dimension
in the monomial equivalence class of $\mathcal C^{(\ell)}$
when $q$ is even, and the maximal Hermitian hull dimension
in that of $\mathcal C_{q^2}^{(\ell)}$
for every prime power $q$.
In odd characteristic, Euclidean scaling restricts the
normalized weights and $\tau$ to nonzero squares, giving
a different rank minimization problem.

We next show that every hull dimension between zero and
the fixed-$\tau$ maximum can be attained by rescaling
only the evaluation coordinates, while keeping the
multipliers of the two appended coordinates fixed.

\begin{corollary}\label{cor:fixed-tau-attainable-hulls}
Let $k\ge4$, $n>k$, and let
$\alpha_1,\ldots,\alpha_n\in\mathbb F_q$
be pairwise distinct. Fix
$\eta\in\mathbb F_q^*$,
$\delta\in\mathbb F_q$, $\ell\in\{k-2,k-1\}$,
and $\tau\in\mathbb F_q^*$.
In either setting of
Proposition~\ref{prop:restricted-scaling-maximal-hull}, fix $a,b$
realizing $\tau$. As $\boldsymbol v$ varies, the hull dimensions
are exactly
\[
\{0,1,\ldots,h^{(\ell)}(\tau)\}.
\]
\end{corollary}

\begin{proof}
Fix $a,b$ realizing $\tau$, and choose the evaluation-coordinate
multipliers attaining $h^{(\ell)}(\tau)$.
Denote the resulting code by $C_0$.

Since evaluation on $\mathcal A$ is injective on
$S_{k,\ell,\eta}$, the evaluation coordinates of $C_0$
contain an information set of size $k$.
After grouping these coordinates first and applying
invertible row operations, a generator matrix of $C_0$
has the form
\[
G=[I_k\mid A].
\]

We show that rescaling only these $k$ evaluation
coordinates can make the corresponding generator
Gram matrix nonsingular.
For nonzero scalars $s_1,\ldots,s_k$, the rescaled
generator matrix is
\[
G_{\boldsymbol s}
=
[\operatorname{diag}(s_1,\ldots,s_k)\mid A].
\]
Its Euclidean and Hermitian Gram matrices are,
respectively,
\[
\begin{aligned}
G_{\boldsymbol s}G_{\boldsymbol s}^{\mathsf T}
&=AA^{\mathsf T}
+\operatorname{diag}(s_1^2,\ldots,s_k^2),\\
G_{\boldsymbol s}G_{\boldsymbol s}^{\dagger}
&=AA^\dagger
+\operatorname{diag}(N(s_1),\ldots,N(s_k)).
\end{aligned}
\]
Since the square map in even characteristic and
the norm map in the Hermitian setting are surjective,
it suffices to choose $d_1,\ldots,d_k\in\mathbb F_q^*$
such that
\[
\det\bigl(M+\operatorname{diag}(d_1,\ldots,d_k)\bigr)
\ne0,
\]
where $M=AA^{\mathsf T}$ in the Euclidean case and
$M=AA^\dagger$ in the Hermitian case.

Choose $d_k,d_{k-1},\ldots,d_1$ successively so that
each trailing principal submatrix of
$M+\operatorname{diag}(d_1,\ldots,d_k)$ is nonsingular.
At each step, the determinant is linear in the new
variable $d_j$, with nonzero coefficient equal to
the determinant of the previously chosen principal
submatrix.
Thus at most one value of $d_j$ is excluded.
Since $q\ge n>k\ge4$, there are at least two
nonzero choices for each $d_j$.
Hence the required diagonal entries exist,
and the resulting code is LCD.

Finally, perform the chosen coordinate rescalings
one at a time, starting from $C_0$.
Each rescaling changes the generator Gram matrix
by a matrix of rank at most one.
Consequently, the hull dimensions of consecutive
codes differ by at most one.
Since the initial hull dimension is
$h^{(\ell)}(\tau)$ and the final hull dimension
is zero, every intermediate integer dimension
is attained.
Throughout these rescalings, the multipliers
$a,b$ of the two appended coordinates remain fixed.
\end{proof}

\section{Moment representations and rank-one perturbations}\label{sec:hankel}

We first characterize the moment sequences representable
by arbitrary weights on the fixed evaluation set $\mathcal A$.
We then address the requirement that all weights be nonzero,
using rank-one perturbations when necessary.

Fix $\ell\in\{k-2,k-1\}$ and $\tau\in\mathbb F_q^*$.
Since the dual code has dimension $r+2$,
Proposition~\ref{prop:restricted-scaling-maximal-hull}
and the hull-dimension identity in
Subsection~\ref{sec:gram-conventions} give
\begin{equation}\label{eq:fixed-tau-hull-maximum}
h^{(\ell)}(\tau)
=r+2-\min_{\boldsymbol\rho\in(\mathbb F_q^*)^n}
\rank\left(
\left(\sum_{i=1}^n\rho_i\alpha_i^{s+t}\right)_{0\le s,t\le r+1}
+B^{(\ell)}(\tau)
\right).
\end{equation}
A matrix is \emph{Hankel} if its entries are constant along each
anti-diagonal. In~\eqref{eq:fixed-tau-hull-maximum} the evaluation
block has this property because its $(s,t)$-entry depends only on
$s+t$. For weights $\boldsymbol\rho\in\mathbb F_q^n$, its entries
are given by the \emph{weighted moments} on $\mathcal A$,
\begin{equation}\label{eq:weighted-power-moments}
\mu_j(\boldsymbol\rho)
:=\sum_{i=1}^n\rho_i\alpha_i^j,
\qquad j\ge0.
\end{equation}
We use \emph{moment} as shorthand for these weighted power sums.
For $m\ge1$ and an arbitrary sequence
$\boldsymbol t=(t_0,\ldots,t_{2m-2})$ over a field $\mathbb K$,
write
\begin{equation}\label{eq:hankel-matrix-definition}
H_m(\boldsymbol t)=(t_{i+j})_{0\le i,j<m}.
\end{equation}
The space of all $m\times m$ Hankel matrices is
$\mathcal H_m(\mathbb K)=\{H_m(\boldsymbol t):
\boldsymbol t\in\mathbb K^{2m-1}\}$.
For a weight vector, let
$\boldsymbol\mu(\boldsymbol\rho)
=(\mu_0(\boldsymbol\rho),\ldots,\mu_{2m-2}(\boldsymbol\rho))$.
Thus the evaluation block in~\eqref{eq:fixed-tau-hull-maximum}
is $H_{r+2}(\boldsymbol\mu(\boldsymbol\rho))$.

For a prescribed sequence
$\boldsymbol\mu\in\mathbb F_q^{2m-1}$,
we first determine whether there exists
$\boldsymbol\rho\in\mathbb F_q^n$
such that
$\boldsymbol\mu=\boldsymbol\mu(\boldsymbol\rho)$,
and then whether such a representation exists
with $\boldsymbol\rho\in(\mathbb F_q^*)^n$.

\begin{proposition}
\label{prop:realizable-moment-recurrence}
Let $m\ge1$.
The map $\boldsymbol\rho\mapsto\boldsymbol\mu(\boldsymbol\rho)$
from $\mathbb F_q^n$ to $\mathbb F_q^{2m-1}$ has rank
$\min\{n,2m-1\}$.
In particular, it is surjective if $2m-1\le n$.
If $2m-1>n$, a moment sequence
$(\mu_0,\ldots,\mu_{2m-2})\in\mathbb F_q^{2m-1}$ belongs to
its image if and only if
\begin{equation}\label{eq:dual-moment-recurrence}
\mu_{n+j}+\sigma_1\mu_{n+j-1}+\cdots+\sigma_n\mu_j=0,
\qquad
0\le j\le2m-2-n.
\end{equation}
These $2m-1-n$ relations are independent.
\end{proposition}

\begin{proof}
The coefficient matrix
$(\alpha_i^j)_{\substack{0\le j\le2m-2\\1\le i\le n}}$
is a Vandermonde matrix of rank $\min\{n,2m-1\}$,
which proves the rank and surjectivity assertions. If $2m-1>n$,
each represented sequence satisfies~\eqref{eq:dual-moment-recurrence}
because $Q(\alpha_i)=0$ for every $i$. Conversely, the first $n$
moments determine unique weights by an invertible Vandermonde system.
The recurrence forces the later moments of these weights
to coincide with the prescribed ones. The recurrence
equations are independent because their coefficients on
$\mu_n,\ldots,\mu_{2m-2}$ form a lower triangular matrix with diagonal
entries one.
\end{proof}

If fewer than $n$ initial moments are prescribed, the kernel of the
moment map gives some freedom in the choice of weights. The next lemma
uses this freedom to make every coordinate nonzero when $q>n$.

\begin{lemma}\label{lem:weighted-moment}
Let $\alpha_1,\ldots,\alpha_n\in\mathbb F_q$ be pairwise distinct,
let $0\le d\le n-2$, and assume $q>n$. For any
$\mu_0,\ldots,\mu_d\in\mathbb F_q$, there exist
$\rho_1,\ldots,\rho_n\in\mathbb F_q^*$ such that
\[
\sum_{i=1}^n\rho_i\alpha_i^j=\mu_j,
\qquad 0\le j\le d.
\]
\end{lemma}

\begin{proof}
The Vandermonde matrix $V_d=(\alpha_i^j)_{0\le j\le d,\ 1\le i\le n}$
has full row rank, so the equations have a solution
$\boldsymbol\rho^{(0)}$. Lemma~\ref{lem:lagrange-moments} gives
$\boldsymbol u=(u_1,\ldots,u_n)$ with
$V_d\boldsymbol u^{\mathsf T}=0$ and every $u_i\ne0$.
Thus $\boldsymbol\rho^{(0)}+c\boldsymbol u$ remains a solution.
Each coordinate excludes one value of $c$, and $q>n$ leaves a
choice for which all coordinates are nonzero.
\end{proof}

The preceding lemma applies when the prescribed moments
leave freedom to adjust the weights.
When the first $n$ moments are prescribed, however,
the weights are uniquely determined and may contain zeros.
Moreover, prescribing one additional moment may violate
the required recurrence relation.
The following lemma addresses these difficulties by
perturbing the moments from $t_s$ to $t_s+c\beta^s$.
As explained after its proof, this perturbation changes
the corresponding matrices by a matrix of rank at most one.

\begin{lemma}\label{lem:moment-perturbation}
Let $t_0,\ldots,t_{n-1}\in\mathbb F_q$, and let
$\boldsymbol\rho^{(0)}\in\mathbb F_q^n$ be the unique solution of
\[
\sum_{i=1}^n\rho_i^{(0)}\alpha_i^s=t_s,
\qquad 0\le s<n.
\]
For $\beta\in\mathbb F_q\setminus\mathcal A$ and
$c\in\mathbb F_q^*$, define
\[
\rho_i=\rho_i^{(0)}+c\,u_i\frac{Q(\beta)}{\beta-\alpha_i},
\qquad 1\le i\le n.
\]
These weights satisfy
\[
\sum_{i=1}^n\rho_i\alpha_i^s=t_s+c\beta^s,
\qquad 0\le s<n.
\]
\begin{enumerate}[label=\textup{(\roman*)}]
\item If $q>n$ and $\rho_j^{(0)}=0$ for some $j$, then
$\beta$ and $c$ can be chosen so that every $\rho_i$ is nonzero.
\item Let $t_n\in\mathbb F_q$ and suppose that
\[
R=t_n+\sigma_1t_{n-1}+\cdots+\sigma_nt_0\ne0.
\]
If $\beta$ satisfies
\begin{equation}\label{eq:moment-perturbation-conditions}
\beta\ne\alpha_i,\qquad
\rho_i^{(0)}(\beta-\alpha_i)-Ru_i\ne0,
\qquad 1\le i\le n,
\end{equation}
then taking $c=-R/Q(\beta)$ makes every $\rho_i$ nonzero,
and the preceding moment identity also holds for $s=n$.
Such a $\beta$ exists whenever $q>2n$.
\end{enumerate}
\end{lemma}

\begin{proof}
Put $\omega_i(\beta)=u_iQ(\beta)/(\beta-\alpha_i)$.
Lagrange interpolation, applied to $x^s$ for $0\le s<n$
and to $x^n-Q(x)$, gives
\begin{equation}\label{eq:external-evaluation-moments}
\sum_i\omega_i(\beta)\alpha_i^s=\beta^s\quad\text{for }0\le s<n,
\qquad
\sum_i\omega_i(\beta)\alpha_i^n=\beta^n-Q(\beta).
\end{equation}
The first identity proves the formula for the perturbed moments.

For part~\textup{(i)}, choose any
$\beta\in\mathbb F_q\setminus\mathcal A$, which is possible since
$q>n$. Every $\omega_i(\beta)$ is nonzero.
A zero coordinate of $\boldsymbol\rho^{(0)}$ becomes nonzero for
every $c\ne0$, while each nonzero coordinate excludes one value
of $c$. Thus at most $n-1$ nonzero values are excluded, and
$q-1\ge n$ gives the required choice.

For part~\textup{(ii)}, taking $c=-R/Q(\beta)$ gives
$\rho_i=\rho_i^{(0)}-Ru_i/(\beta-\alpha_i)$, which is nonzero
by~\eqref{eq:moment-perturbation-conditions}.
The original weights have degree-$n$ moment $t_n-R$ by the
recurrence associated with $Q$.
Since $cQ(\beta)=-R$,
\[
\sum_i\rho_i\alpha_i^n
=(t_n-R)+c\bigl(\beta^n-Q(\beta)\bigr)
=t_n+c\beta^n.
\]
The conditions exclude the $n$ points of $\mathcal A$ and
the roots of $\rho_i^{(0)}(z-\alpha_i)-Ru_i$ for $1\le i\le n$.
Each of these polynomials is nonzero and has degree at most one,
so at most $2n$ values of $\beta$ are excluded.
\end{proof}

Let $f=(f_1,\ldots,f_d)^{\mathsf T}\in\mathbb F_q[x]^d$ be
the polynomial column defining the evaluation coordinates,
and suppose that the prescribed moments include every degree
occurring in the products $f_if_j$.
The $(i,j)$ matrix entry contributed by these coordinates is
obtained by expanding $f_i(x)f_j(x)$ and replacing each $x^s$
with $t_s$.
Thus replacing $t_s$ by $t_s+c\beta^s$ adds
$cf_i(\beta)f_j(\beta)$ to this entry.
With the terms from the appended coordinates fixed, the matrix
therefore changes by
\[
c\,f(\beta)f(\beta)^{\mathsf T},
\]
so its rank changes by at most one.

For Hankel matrices, take
$f(z)=\boldsymbol h_m(z):=(1,z,\ldots,z^{m-1})^{\mathsf T}$.
The same argument applies to the generator Gram matrices in
Section~\ref{sec:generator}.

\section{\texorpdfstring{Dual Gram analysis for $2\le r\le k-2$}{Dual Gram analysis for 2 <= r <= k-2}}\label{sec:dual-maxima}

For $2\le r\le k-2$, the dual code has dimension $r+2\le k$,
so we use the dual Gram matrix.
By~\eqref{eq:fixed-tau-hull-maximum}, the fixed-$\tau$ hull
problem is equivalent to minimizing the rank of the dual Gram matrix over
nonzero evaluation weights.
We first determine the minimum over the full Hankel space,
then examine whether it is attained by nonzero weights on
$\mathcal A$.

Throughout this section, let $k\ge4$ and $2\le r=n-k\le k-2$,
and take $\ell=k-2$.
Fix $\eta\in\mathbb F_q^*$ and $\delta\in\mathbb F_q$, together
with pairwise distinct evaluation points
$\alpha_1,\ldots,\alpha_n\in\mathbb F_q$.
For $\tau\in\mathbb F_q^*$, put
\[
B(\tau)=F_{k-2}\operatorname{diag}(1,\tau)F_{k-2}^{\mathsf T},
\qquad D=\delta^2+\tau,
\qquad \Lambda_2=\sigma_1^2-\sigma_2.
\]

The variable Hankel matrix in
\eqref{eq:fixed-tau-hull-maximum} has order $r+2$
and is determined by the $2r+3$ moments
$\mu_0,\ldots,\mu_{2r+2}$.
These moments arise from the $n=k+r$ evaluation weights
$\rho_1,\ldots,\rho_n$.
The relevant linear map is
\[
\boldsymbol\rho\longmapsto
\boldsymbol\mu(\boldsymbol\rho)
=
\bigl(\mu_0(\boldsymbol\rho),\ldots,
\mu_{2r+2}(\boldsymbol\rho)\bigr).
\]
By Proposition~\ref{prop:realizable-moment-recurrence},
the image of this linear map has codimension
\begin{equation}\label{eq:moment-codimension}
(2r+3)-\min\{k+r,2r+3\}
=\max\{0,r-k+3\}.
\end{equation}

The three parameter ranges in Table~\ref{tab:moment-ranges}
differ in whether the moments can be prescribed arbitrarily,
whether their representing weights are unique,
and whether a recurrence constraint must be satisfied.
\begin{center}
\begin{minipage}{\linewidth}
\small
\captionsetup{hypcap=false,font=small,labelfont=bf,
  justification=raggedright,singlelinecheck=false}
\captionof{table}{Mapping properties of the linear map $\boldsymbol\rho\mapsto\boldsymbol\mu(\boldsymbol\rho)$ for $2\le r\le k-2$.}
\label{tab:moment-ranges}
\setlength{\tabcolsep}{5pt}
\renewcommand{\arraystretch}{1.25}
\begin{tabular*}{\linewidth}{@{\extracolsep{\fill}}
  >{\raggedright\arraybackslash}p{0.19\linewidth}
  >{\raggedright\arraybackslash}p{0.34\linewidth}
  >{\raggedright\arraybackslash}p{0.39\linewidth}@{}}
\toprule
Range & Mapping properties & Representing weights\\
\midrule
$2\le r\le k-4$ & Surjective with a nontrivial kernel
& Nonunique.\\
$r=k-3$ & Bijective
& Unique.\\
$r=k-2$ & Injective with image of codimension one
& Unique when the recurrence condition holds.\\
\bottomrule
\end{tabular*}
\par\smallskip
\noindent\emph{Note.} Zero weights are allowed in this table.
For $2\le r\le k-4$ and $q>n$, Lemma~\ref{lem:weighted-moment}
gives a representation with every weight nonzero.
For $r=k-3$ and $r=k-2$, the unique representing weights
must be checked for zero entries.
\end{minipage}
\end{center}

\subsection{Rank minimization over the Hankel space}\label{sec:corner-ranks}

Rank distributions of Hankel matrices over finite fields have
been studied in~\cite{dwivedi2022hankel}.
We study the minimum of $\rank(M+B)$ over Hankel matrices $M$,
where $B$ is a fixed symmetric matrix vanishing outside its
lower-right $3\times3$ principal block.
The following theorem holds over an arbitrary field.

\begin{theorem}\label{thm:hankel-distance-three-row}
Let $r\ge2$ and let $\mathbb K$ be a field. Let $B$ be a symmetric
$(r+2)\times(r+2)$ matrix over $\mathbb K$, with rows and columns indexed
by $0,\ldots,r+1$, such that $B_{ij}=0$ unless
$i,j\in\{r-1,r,r+1\}$. Write its lower-right $3\times3$ principal block
as $B_0=(b_{ij})_{0\le i,j\le2}$.
\begin{enumerate}[label=\textup{(\roman*)}]
\item If $b_{00}\ne0$, then
\[
\min_{M\in\mathcal H_{r+2}(\mathbb K)}\rank(M+B)=1.
\]
This minimum is attained by $M=H_{r+2}(\boldsymbol\mu)$ with
\[
\mu_0=\cdots=\mu_{2r-1}=0,\qquad
(\mu_{2r},\mu_{2r+1},\mu_{2r+2})=(\xi,\upsilon,\zeta),
\]
where
\begin{equation}\label{eq:rank-one-tail}
\begin{aligned}
\xi&=b_{01}^2b_{00}^{-1}-b_{11},\\
\upsilon&=b_{01}(b_{02}+\xi)b_{00}^{-1}-b_{12},\\
\zeta&=(b_{02}+\xi)^2b_{00}^{-1}-b_{22}.
\end{aligned}
\end{equation}
For $r\ge3$, this is the unique Hankel matrix $M$
for which $\rank(M+B)=1$.
\item If $b_{00}=0$ and $b_{01}\ne0$, then
\[
\min_{M\in\mathcal H_{r+2}(\mathbb K)}\rank(M+B)
=
\begin{cases}
1,&r=2,\\
2,&r\ge3.
\end{cases}
\]
For $r\ge3$, rank two is attained by every moment vector satisfying
\begin{equation}\label{eq:rank-two-prefix}
\mu_0=\cdots=\mu_{2r-2}=0,\qquad
\mu_{2r-1}=-b_{01},\qquad
\mu_{2r}=-b_{11},
\end{equation}
where $\mu_{2r+1},\mu_{2r+2}\in\mathbb K$ are arbitrary.
\end{enumerate}
\end{theorem}

\begin{proof}
Rank zero is impossible in either case, since it would make $B$
Hankel. If $b_{00}\ne0$, the entries
$B_{r-1,r-1}=b_{00}$ and $B_{r-2,r}=0$ differ on the same
anti-diagonal. If $b_{00}=0$ and $b_{01}\ne0$, the entries
$B_{r-1,r}=b_{01}$ and $B_{r-2,r+1}=0$ give the same obstruction.

Suppose $r\ge3$ and $G=H_{r+2}(\boldsymbol\mu)+B$ has rank one.
Some diagonal entry $G_{aa}$ is nonzero, since otherwise its
principal $2\times2$ minors would give $G_{ij}^2=0$ for all $i,j$.
All columns are multiples of column $a$, so symmetry gives
$G=\gamma\boldsymbol z\boldsymbol z^{\mathsf T}$ with
$\boldsymbol z=G_{\cdot a}$ and $\gamma=G_{aa}^{-1}$.
For $0\le i\le r-3$ and $0\le j\le r$, both entries
$(i,j+1)$ and $(i+1,j)$ lie outside the support of $B$.
The Hankel identities therefore give
\[
z_i z_{j+1}=z_{i+1}z_j.
\]
If some $z_i\ne0$ with $i\le r-3$, these identities imply
$z_{j+1}=(z_{i+1}/z_i)z_j$ for every $0\le j\le r$.
Then $G$ is Hankel, contradicting the fact that
$B=G-H_{r+2}(\boldsymbol\mu)$ is not Hankel.
Thus $z_0=\cdots=z_{r-3}=0$.
Taking $i=r-3$ and $j=r-2$ also gives $z_{r-2}^2=0$.
The zero rows indexed by $0,\ldots,r-2$ now imply
\[
\mu_0=\cdots=\mu_{2r-1}=0.
\]

Consequently, $G_{r-1,r-1}=b_{00}$ and
$G_{r-1,r}=b_{01}$.
If $b_{00}=0$, a zero diagonal entry of a symmetric rank-one
matrix forces its entire row to vanish, contradicting
$b_{01}\ne0$.
If $b_{00}\ne0$, using $G_{r-1,r-1}$ as a pivot forces
the entries $(r,r)$, $(r,r+1)$, and $(r+1,r+1)$,
and hence the moments in~\eqref{eq:rank-one-tail}.
This proves uniqueness in part~\textup{(i)} for $r\ge3$
and excludes rank one in part~\textup{(ii)} for $r\ge3$.

Conversely, if $b_{00}\ne0$, the moments in part~\textup{(i)}
make $H_{r+2}(\boldsymbol\mu)+B$ zero outside its lower-right
$3\times3$ block, which equals
\[
\frac1{b_{00}}
\begin{pmatrix}b_{00}\\b_{01}\\b_{02}+\xi\end{pmatrix}
\begin{pmatrix}b_{00}&b_{01}&b_{02}+\xi\end{pmatrix}.
\]
Thus rank one is attained for every $r\ge2$.

Now let $b_{00}=0$, $b_{01}\ne0$, and $r\ge3$.
Take the moments in~\eqref{eq:rank-two-prefix}, and write
$Y=\mu_{2r+1}$ and $Z=\mu_{2r+2}$.
The resulting matrix is zero outside the coordinates
$r-2,r-1,r,r+1$, on which it has the block
\begin{equation}\label{eq:rank-two-exceptional-block}
\begin{pmatrix}
0&0&0&-b_{01}\\
0&0&0&b_{02}-b_{11}\\
0&0&0&Y+b_{12}\\
-b_{01}&b_{02}-b_{11}&Y+b_{12}&Z+b_{22}
\end{pmatrix}.
\end{equation}
The first three rows are scalar multiples of $(0,0,0,1)$,
and the principal minor on the first and fourth coordinates
has determinant $-b_{01}^2\ne0$.
Thus its rank is two for every $Y,Z$.

Finally, let $r=2$, $b_{00}=0$, and $b_{01}\ne0$, and put
\[
\theta=\frac{b_{11}-b_{02}}{b_{01}},\qquad
\boldsymbol z=(1,\theta,\theta^2,\theta^3-b_{01})^{\mathsf T}.
\]
The symmetric matrix $M=\boldsymbol z\boldsymbol z^{\mathsf T}-B$
satisfies
$M_{02}=M_{11}$, $M_{03}=M_{12}$, and $M_{13}=M_{22}$,
so it is Hankel.
Since $z_0=1$, the matrix
$M+B=\boldsymbol z\boldsymbol z^{\mathsf T}$ has rank one.
\end{proof}

For $r=2$, rank-one solutions may have $\mu_0\ne0$.
Appendix~\ref{app:k5-r2} parametrizes the rank-one
moment vectors needed for the case $(k,r)=(5,2)$.

We now apply Theorem~\ref{thm:hankel-distance-three-row}
to the matrix $B(\tau)$ associated with the TRL code.
The entries of its lower-right $3\times3$ principal block are
\begin{equation}\label{eq:leading-block-entries}
\begin{aligned}
b_{00}&=\eta^2D,\\
b_{01}&=-\eta\delta-\eta^2\sigma_1D,\\
b_{02}&=\eta\delta\sigma_1+\eta(1+\eta\Lambda_2)D,\\
b_{11}&=1+2\eta\delta\sigma_1+\eta^2\sigma_1^2D,\\
b_{12}&=-\sigma_1-\delta(1+\eta\Lambda_2)-\eta\delta\sigma_1^2
        -\eta\sigma_1(1+\eta\Lambda_2)D,\\
b_{22}&=\sigma_1^2+2\delta\sigma_1(1+\eta\Lambda_2)
        +(1+\eta\Lambda_2)^2D.
\end{aligned}
\end{equation}
For $D\ne0$, Theorem~\ref{thm:hankel-distance-three-row}(i)
applies because $b_{00}=\eta^2D\ne0$.
For $D=0$, the condition $\tau=-\delta^2\ne0$ gives
$b_{01}=-\eta\delta\ne0$, so part~\textup{(ii)} applies.

The next three subsections determine the minimum rank of the dual
Gram matrix for nonzero weights on $\mathcal A$ in the ranges
of Table~\ref{tab:moment-ranges}.

\subsection{\texorpdfstring{Surjectivity of the linear map: $2\le r\le k-4$}{Surjectivity of the linear map: 2 <= r <= k-4}}\label{sec:spectra}

For $2\le r\le k-4$, the linear map
$\boldsymbol\rho\mapsto\boldsymbol\mu(\boldsymbol\rho)$
is surjective and has a nontrivial kernel.
If $q>n$, Lemma~\ref{lem:weighted-moment} shows that every
prescribed moment sequence can be represented by nonzero weights.

\begin{theorem}\label{thm:free-maximum-hulls}
Suppose $2\le r\le k-4$ and $q>n$. Then
\[
h^{(k-2)}(\tau)=
\begin{cases}
r+1,&r=2\ \text{or}\ D\ne0,\\
r,&r\ge3\ \text{and}\ D=0.
\end{cases}
\]
\end{theorem}

\begin{proof}
Since $2r+2\le n-2$, Lemma~\ref{lem:weighted-moment} provides
nonzero weights for every sequence $(\mu_0,\ldots,\mu_{2r+2})$.
Thus every Hankel matrix in $\mathcal H_{r+2}(\mathbb F_q)$
can be represented by such weights.
If $D\ne0$, Theorem~\ref{thm:hankel-distance-three-row}(i)
shows that the minimum rank of the dual Gram matrix is one.
If $D=0$, then $\tau=-\delta^2\ne0$, so
$b_{01}\ne0$. Part~\textup{(ii)} of the same theorem gives minimum
rank one for $r=2$ and two for $r\ge3$.
Equation~\eqref{eq:fixed-tau-hull-maximum} gives the result.
\end{proof}

Under the same hypotheses, maximizing over
$\tau\in\mathbb F_q^*$ yields
$h_{\max}^{(k-2)}=r+1$.
Indeed, since $q>n$, we can choose
$\tau\in\mathbb F_q\setminus\{0,-\delta^2\}$,
for which $D\ne0$.

\subsection{\texorpdfstring{Unique moment representations: $r=k-3$}{Unique moment representations: r=k-3}}
\label{sec:beyond}

Assume $r=k-3\ge3$, so $n=2r+3$ and the dual Gram matrix uses
exactly the moments $\mu_0,\ldots,\mu_{n-1}$.
The nonsingular Vandermonde system gives a unique representation
of every such sequence, with no recurrence constraint.
For $D\ne0$, Theorem~\ref{thm:hankel-distance-three-row}(i)
also gives a unique moment sequence $\boldsymbol\mu$ for which
$H_{r+2}(\boldsymbol\mu)+B(\tau)$ has rank one.
It remains to determine whether its representing weights are nonzero.

By Theorem~\ref{thm:hankel-distance-three-row}(i)
and the Lagrange identities in
Lemma~\ref{lem:lagrange-moments},
the moment sequence $\boldsymbol\mu$ has unique
representing weights $\rho_i=u_iP_\tau(\alpha_i)$,
where $P_\tau$ is a quadratic polynomial.
Substituting the prescribed moments gives the following
explicit expressions. Define
\begin{equation}\label{eq:compatibility-polynomial}
\Phi(D)=\eta(\eta\Lambda_2+2)D^2
+(\delta\eta\sigma_1-1)D+\delta^2.
\end{equation}
For $D\ne0$, put
\begin{equation}\label{eq:critical-polynomial}
P_\tau(z)=-\frac{\tau}{\eta^2D^3}
\left[
\eta^2D^2z^2
-\eta D(\delta+\eta\sigma_1D)z
+\Phi(D)
\right].
\end{equation}
Since $u_i\ne0$, the representing weights
$\rho_i=u_iP_\tau(\alpha_i)$ are all nonzero
if and only if $P_\tau$ has no zeros in $\mathcal A$.

\begin{theorem}\label{thm:critical-lower-exact-rank}
Suppose $r=k-3\ge3$ and $q>n$. Then
\[
h^{(k-2)}(\tau)=
\begin{cases}
r+1,&D\ne0\ \text{and}\ P_\tau(\alpha_i)\ne0\ \text{for all }i,\\
r,&\text{otherwise}.
\end{cases}
\]
\end{theorem}

\begin{proof}
For $D\ne0$, the sequence specified by
Theorem~\ref{thm:hankel-distance-three-row}(i) satisfies
$\mu_0=\cdots=\mu_{2r-1}=0$ and
$(\mu_{2r},\mu_{2r+1},\mu_{2r+2})=(\xi,\upsilon,\zeta)$.
Substituting~\eqref{eq:leading-block-entries}
into~\eqref{eq:rank-one-tail} gives
\begin{equation}\label{eq:explicit-terminal-moments}
\begin{aligned}
\xi&=-\frac{\tau}{D},\\
\upsilon+\sigma_1\xi
 &=\frac{\tau}{\eta D^2}(\delta+\eta\sigma_1D),\\
\zeta+\sigma_1\upsilon+\sigma_2\xi
 &=-\frac{\tau}{\eta^2D^3}\Phi(D).
\end{aligned}
\end{equation}
Hence
\[
P_\tau(z)=\xi z^2+(\upsilon+\sigma_1\xi)z
 +\zeta+\sigma_1\upsilon+\sigma_2\xi.
\]
Since $n=2r+3$, Lemma~\ref{lem:lagrange-moments} shows that
$\rho_i=u_iP_\tau(\alpha_i)$ represent all $n$ prescribed moments.
The nonsingular Vandermonde system makes these weights unique.
As $u_i\ne0$, rank one is attainable with nonzero weights exactly
when $P_\tau$ has no zero in $\mathcal A$.

If $P_\tau$ has a zero in $\mathcal A$, this moment sequence has
no representation with nonzero weights.
Lemma~\ref{lem:moment-perturbation}(i) gives nonzero weights whose
Gram matrix differs from the corresponding rank-one matrix by
a matrix of rank at most one.
The minimum rank of the dual Gram matrix is therefore two, since rank zero
is excluded by Theorem~\ref{thm:hankel-distance-three-row}.

For $D=0$, Theorem~\ref{thm:hankel-distance-three-row}(ii)
gives a lower bound of two for the rank of the dual Gram matrix.
Apply Lemma~\ref{lem:weighted-moment} with $d=2r=n-3$
to the moments in~\eqref{eq:rank-two-prefix}.
The resulting weights are nonzero, and every choice of the remaining
moments gives a dual Gram matrix of rank two by the same theorem.
Thus these weights attain the lower bound.
Equation~\eqref{eq:fixed-tau-hull-maximum} gives the stated
hull dimensions.
\end{proof}

\begin{corollary}\label{cor:full-max-kminus3}
Assume $r=k-3\ge3$ and $q>n$.
Then $h_{\max}^{(k-2)}=r+1$ if and only if some
$\tau\in\mathbb F_q^*$ satisfies
$D\ne0$ and $P_\tau(\alpha_i)\ne0$ for every $i$.
Otherwise, $h_{\max}^{(k-2)}=r$.
In particular, $h_{\max}^{(k-2)}=r+1$ whenever $q>2n+2$.
\end{corollary}

\begin{proof}
The first assertion follows by maximizing
Theorem~\ref{thm:critical-lower-exact-rank} over $\tau$.
For the sufficient field-size condition, regard $D$ as the variable
and recover $\tau=D-\delta^2$.
We must avoid $D=0$ and $D=\delta^2$, as well as all zeros of the
representing weights.
For each $z=\alpha_i$, the bracket
in~\eqref{eq:critical-polynomial} is a nonzero polynomial in $D$
of degree at most two.
Its constant term is $\delta^2$ when $\delta\ne0$, and its
coefficient of $D$ is $-1$ when $\delta=0$.
Its roots, together with the inadmissible values $0$ and $\delta^2$,
exclude at most $2n+2$ values of $D$.
Thus $q>2n+2$ leaves a value giving nonzero weights and hull
dimension $r+1$.
\end{proof}

\subsection{\texorpdfstring{One recurrence constraint: $r=k-2$}{One recurrence constraint: r=k-2}}
\label{sec:dual-one-recurrence}

Assume $r=k-2\ge3$, so $n=2r+2$ and the code has length $2k$.
The dual Gram matrix now involves $n+1$ moments but only $n$
weights.
The first $n$ moments determine unique weights, and
Proposition~\ref{prop:realizable-moment-recurrence} shows that the
last moment is represented exactly when the single recurrence
in~\eqref{eq:dual-moment-recurrence} holds.
Attaining rank one for the dual Gram matrix therefore requires
both the recurrence relation and nonzero representing weights.

By Theorem~\ref{thm:hankel-distance-three-row}, the dual Gram matrix
has rank at least one for every $\tau\in\mathbb F_q^*$.
Since $r+2=k$, this yields
\begin{equation}\label{eq:balanced-hull-bound}
h^{(k-2)}(\tau)\le k-1.
\end{equation}

It remains to determine whether hull dimension
$k-1=r+1$ is attainable.
For $D\ne0$, Theorem~\ref{thm:hankel-distance-three-row}(i)
gives the unique moment sequence $\boldsymbol\mu$
for which $H_{r+2}(\boldsymbol\mu)+B(\tau)$ has rank one.
For $r=k-3$, the three terminal moments
$\xi,\upsilon,\zeta$ determine the quadratic
polynomial $P_\tau$.
Here the first $n$ moments end with $\xi,\upsilon$
and determine the linear polynomial $P_D$,
while the remaining moment $\zeta$ enters
the recurrence residual $R_D$.
By the Lagrange identities in
Lemma~\ref{lem:lagrange-moments}, the unique weights
representing the first $n$ moments of this sequence
are $\rho_i=u_iP_D(\alpha_i)$, where
\begin{equation}\label{eq:linear-target-polynomial}
P_D(z)=\frac{\tau}{\eta D^2}
[\delta+\eta(\sigma_1-z)D].
\end{equation}
To express the nonzero-weight condition, put
\[
L_i(D)=\delta+\eta(\sigma_1-\alpha_i)D,
\qquad 1\le i\le n.
\]
Since $P_D(\alpha_i)=\tau L_i(D)/(\eta D^2)$
and $u_i\ne0$, all the representing weights
are nonzero if and only if $L_i(D)\ne0$
for every $i$.

The remaining moment must also satisfy the recurrence
relation in~\eqref{eq:dual-moment-recurrence}.
Substituting $\boldsymbol\mu$ into this relation gives
the residual
\begin{equation}\label{eq:kminus2-residual}
R_D=-\frac{\tau}{\eta^2D^3}\Phi(D),
\end{equation}
where $\Phi(D)$ is defined in
\eqref{eq:compatibility-polynomial}.
Thus the recurrence holds if and only if
$\Phi(D)=0$.

The following theorem characterizes when hull dimension
$r+1$ is attainable and gives conditions for attaining
hull dimension $r$ when the criterion for attaining
$r+1$ fails.
\begin{theorem}\label{thm:first-constrained-lower}
Assume $r=k-2\ge3$ and fix $\tau\in\mathbb F_q^*$.
\begin{enumerate}[label=\textup{(\roman*)},leftmargin=*,beginpenalty=10000]
\item The equality $h^{(k-2)}(\tau)=r+1$ holds if and only if
\[
D\ne0,\quad\Phi(D)=0,\quad
L_i(D)\ne0\ \text{for all }i.
\]
If this criterion fails, then $h^{(k-2)}(\tau)\le r$.
\item The equality $h^{(k-2)}(\tau)=r$ holds in each of the
following cases:
\begin{enumerate}[label=\textup{(\alph*)},leftmargin=*]
\item $D=0$ and $q>n$.
\item $D\ne0$, $\Phi(D)=0$, and $L_i(D)=0$ for at least one $i$.
\item $D\ne0$, $\Phi(D)\ne0$, and there exists
$\beta\in\mathbb F_q\setminus\mathcal A$ such that
\[
P_D(\alpha_i)(\beta-\alpha_i)-R_D\ne0
\qquad\text{for every }1\le i\le n.
\]
\end{enumerate}
In particular, if $q>2n$, then $h^{(k-2)}(\tau)=r$
whenever the criterion in part~\textup{(i)} fails.
\end{enumerate}
\end{theorem}

\begin{proof}
To prove (i), suppose first that $D\ne0$.
Theorem~\ref{thm:hankel-distance-three-row}(i)
determines the unique sequence
$\boldsymbol\mu=(0,\ldots,0,\xi,\upsilon,\zeta)$ for which
$H_{r+2}(\boldsymbol\mu)+B(\tau)$ has rank one.
By~\eqref{eq:explicit-terminal-moments},
$P_D(z)=\xi z+\upsilon+\sigma_1\xi$ and
$R_D=\zeta+\sigma_1\upsilon+\sigma_2\xi$.
Lemma~\ref{lem:lagrange-moments} and the nonsingularity of the
Vandermonde matrix give $\rho_i^{(0)}=u_iP_D(\alpha_i)$ as the
only weights representing
the first $n$ moments of $\boldsymbol\mu$.
Proposition~\ref{prop:realizable-moment-recurrence} shows that they
represent all its moments if and only if $R_D=0$.
By~\eqref{eq:kminus2-residual}, this is equivalent to $\Phi(D)=0$.
Moreover, $P_D(\alpha_i)=\tau L_i(D)/(\eta D^2)$, so these weights
are all nonzero exactly when every $L_i(D)$ is nonzero.
If $D=0$, Theorem~\ref{thm:hankel-distance-three-row}(ii) excludes
rank one because $b_{01}=-\eta\delta\ne0$ and $r\ge3$.

We turn to (ii). The preceding argument excludes ranks zero and one
in each of the stated cases,
so it suffices to construct nonzero weights giving rank at most two.

If $D=0$ and $q>n$, apply Lemma~\ref{lem:weighted-moment}
with $d=2r=n-2$
to the moments in~\eqref{eq:rank-two-prefix}.
This gives nonzero weights with the prescribed
initial moments.
Every choice of the remaining moments gives a dual Gram matrix
of rank two by
Theorem~\ref{thm:hankel-distance-three-row}(ii).

Next suppose that $D\ne0$, $\Phi(D)=0$, and $L_i(D)=0$ for
some $i$. The weights $u_iP_D(\alpha_i)$ represent all the prescribed
moments, and the corresponding dual Gram matrix has rank one.
The linear coefficient of $P_D$ is
$-\tau/D\ne0$, so exactly one of these weights is zero.
Replacing that zero weight by any nonzero value adds a matrix of
rank one to the dual Gram matrix, giving rank at most two with
all weights nonzero.

Finally, suppose that $D\ne0$, $\Phi(D)\ne0$, and $\beta$ satisfies
the condition in (c). Apply Lemma~\ref{lem:moment-perturbation}(ii)
to the candidate
weights $\rho_i^{(0)}=u_iP_D(\alpha_i)$ and the residual
$R=R_D\ne0$.
The lemma gives the modified weights
\[
\rho_i=u_i\left(P_D(\alpha_i)-\frac{R_D}{\beta-\alpha_i}\right),
\qquad 1\le i\le n.
\]
Since $u_i\ne0$ and $\beta\notin\mathcal A$, these weights are
all nonzero precisely under the stated condition on $\beta$.
They represent the moments $\mu_j+c\beta^j$ for $0\le j\le n$,
where $c=-R_D/Q(\beta)\ne0$.
Since $H_{r+2}(\boldsymbol\mu)+B(\tau)$ has rank one, the
perturbation argument at the end of Section~\ref{sec:hankel}
gives rank at most two for the new dual Gram matrix.

In each case, the preceding lower bound makes the attained rank
exactly two, so $h^{(k-2)}(\tau)=r$.
For $q>2n$, Lemma~\ref{lem:moment-perturbation}(ii) supplies
the required $\beta$ in case~\textup{(c)}.
The same bound implies $q>n$ in case~\textup{(a)}, so the
three cases cover every failure of the criterion in
part~\textup{(i)}.
\end{proof}

For $q>2n$, we may now maximize over $\tau$.
Since $\tau=D-\delta^2$, the maximum is
$h_{\max}^{(k-2)}=r+1$ if and only if some
$D\in\mathbb F_q\setminus\{0,\delta^2\}$ satisfies
$\Phi(D)=0$ and $L_i(D)\ne0$ for every $i$.
Otherwise $h_{\max}^{(k-2)}=r$.

\section{\texorpdfstring{Generator Gram analysis for $r\ge k-1$}{Generator Gram analysis for r >= k-1}}\label{sec:generator}

For $r\ge k-1$, we use the smaller $k\times k$ generator Gram
matrix, whose entries involve moments only through degree $2k$.
Hull dimension $k$ is attained precisely when this matrix vanishes
for nonzero evaluation weights.
We determine the required moments and their representations, then
examine rank one when the zero matrix is unattainable.

Throughout this section, let $k\ge4$ and $r=n-k\ge k-1$, and take
$\ell=k-2$. Fix $\eta\in\mathbb F_q^*$ and $\delta\in\mathbb F_q$
together with pairwise distinct evaluation points
$\alpha_1,\ldots,\alpha_n\in\mathbb F_q$.

\subsection{Normalization and self-orthogonality}\label{sec:generator-gram}

Fix $\tau\in\mathbb F_q^*$.
For the code
$\mathcal C^{(k-2)}\operatorname{diag}(v_1,\ldots,v_n,a,b)$
obtained by scaling the coordinates of $\mathcal C^{(k-2)}$,
dividing its generator Gram matrix by $b^2$
in the Euclidean case or by $N(b)$ in the
Hermitian case gives the contribution of the two appended coordinates
\[
E(D)=\operatorname{diag}\!\left(
\mathbf0_{(k-2)\times(k-2)},
\begin{pmatrix}
1&\delta\\
\delta&D
\end{pmatrix}\right),
\qquad D=\delta^2+\tau.
\]

The corresponding normalized evaluation weights are
$\lambda_i=v_i^2/b^2$ in the Euclidean case and
$\lambda_i=N(v_i)/N(b)$ in the Hermitian case.
In the even-characteristic Euclidean setting
and in the Hermitian setting, the square and norm
maps, respectively, are surjective.
Hence the weights $\lambda_i$ range independently
over $\mathbb F_q^*$.
For the same coordinate scalings, these weights and the normalized
dual weights in Proposition~\ref{prop:restricted-scaling-maximal-hull}
satisfy
\begin{equation}\label{eq:kminus1-gram-correspondence}
\lambda_i\rho_i=\tau u_i^2,\qquad 1\le i\le n.
\end{equation}

Write
\[
g(z)=(1,z,\ldots,z^{k-3},z^{k-2}+\eta z^k,z^{k-1})^{\mathsf T}.
\]
The normalized generator Gram matrix is
\begin{equation}\label{eq:generator-gram}
K_D(\boldsymbol\lambda)=
\sum_{i=1}^n\lambda_i g(\alpha_i)g(\alpha_i)^{\mathsf T}+E(D).
\end{equation}
By the hull-dimension identity in Subsection~\ref{sec:gram-conventions},
\begin{equation}\label{eq:generator-hull-maximum}
h^{(k-2)}(\tau)
=k-\min_{\boldsymbol\lambda\in(\mathbb F_q^*)^n}
\rank K_D(\boldsymbol\lambda).
\end{equation}
For weights $\boldsymbol\lambda\in\mathbb F_q^n$, the moments are
\[
\nu_j=\sum_{i=1}^n\lambda_i\alpha_i^j,
\qquad 0\le j\le2k.
\]
By Proposition~\ref{prop:realizable-moment-recurrence}
with $m=k+1$, the map
$\boldsymbol\lambda\mapsto(\nu_0,\ldots,\nu_{2k})$
has image of codimension
\[
\max\{0,2k+1-n\}=\max\{0,k-r+1\}.
\]
When $n\leq 2k$, the equations defining this image are
\begin{equation}\label{eq:generator-moment-recurrence}
\nu_{n+j}+\sigma_1\nu_{n+j-1}+\cdots+\sigma_n\nu_j=0,
\qquad 0\le j\le2k-n.
\end{equation}
\begin{center}
\begin{minipage}{\linewidth}
\small
\captionsetup{hypcap=false,font=small,labelfont=bf,
  justification=raggedright,singlelinecheck=false}
\captionof{table}{Moment representations for $r\ge k-1$.}
\label{tab:generator-moment-ranges}
\setlength{\tabcolsep}{5pt}
\renewcommand{\arraystretch}{1.25}
\begin{tabular*}{\linewidth}{@{\extracolsep{\fill}}
  >{\raggedright\arraybackslash}p{0.19\linewidth}
  >{\raggedright\arraybackslash}p{0.34\linewidth}
  >{\raggedright\arraybackslash}p{0.39\linewidth}@{}}
\toprule
Range & Mapping properties & Representing weights\\
\midrule
$r=k-1$
& Injective with image of codimension two
& Unique when both recurrence conditions hold.\\
$r=k$
& Injective with image of codimension one
& Unique when the recurrence condition holds.\\
$r=k+1$
& Bijective
& Unique.\\
$r\ge k+2$
& Surjective with a nontrivial kernel
& Nonunique.\\
\bottomrule
\end{tabular*}
\par\smallskip
\noindent\emph{Note.}
Zero weights are allowed in this table.
\end{minipage}
\end{center}
For a prescribed sequence $\boldsymbol s=(s_0,\ldots,s_{2k})$,
write $K_D[\boldsymbol s]$ for the matrix obtained by expanding
$g(z)g(z)^{\mathsf T}$, replacing each $z^j$ by $s_j$ linearly,
and adding $E(D)$. For the weighted moments $\boldsymbol\nu$ of
$\boldsymbol\lambda$, this gives $K_D[\boldsymbol\nu]=K_D(\boldsymbol\lambda)$.
We can now solve for the sequence that makes the matrix zero before
asking whether that sequence has a representation on $\mathcal A$.

\begin{lemma}\label{lem:generator-zero-target}
Let $k\ge4$, $n>k$, and let
$\alpha_1,\ldots,\alpha_n$ be distinct elements of
$\mathbb F_q$. Fix $\eta\in\mathbb F_q^*$,
$\delta\in\mathbb F_q$, and $\tau\in\mathbb F_q^*$,
and put $D=\delta^2+\tau$.
There is a unique sequence $\boldsymbol t$ such that
$K_D[\boldsymbol t]=0$. It is given by
\begin{equation}\label{eq:generator-zero-moments}
\begin{gathered}
t_0=\cdots=t_{2k-3}=0,\qquad t_{2k-2}=-D,\\
t_{2k-1}=-\frac{\delta}{\eta},\qquad
t_{2k}=\frac{2\eta D-1}{\eta^2}.
\end{gathered}
\end{equation}
Thus $K_D(\boldsymbol\lambda)=0$ precisely when the weighted moments
of $\boldsymbol\lambda$ equal $\boldsymbol t$.
\end{lemma}

\begin{proof}
The rows indexed by $1,x,\ldots,x^{k-3}$ give the entries
$t_0,\ldots,t_{2k-6}$ among themselves and $t_{k-1},\ldots,t_{2k-4}$
against the row $x^{k-1}$. These ranges cover $0,\ldots,2k-4$
for $k\ge4$. All these entries must vanish. The entry between
$x^{k-3}$ and $x^{k-2}+\eta x^k$ is $t_{2k-5}+\eta t_{2k-3}$,
so $t_{2k-3}=0$ as well. Every entry between a low-degree row
and the twisted row now vanishes.

The remaining three entries, in the lower-right $2\times2$ block,
vanish precisely when
\[
t_{2k-2}+D=0,\qquad \eta t_{2k-1}+\delta=0,\qquad
2\eta t_{2k-2}+\eta^2t_{2k}+1=0.
\]
Solving gives~\eqref{eq:generator-zero-moments}. These equations
account for every entry of $K_D[\boldsymbol t]$, so the displayed
values also suffice.

\end{proof}

Substituting the moment values in
\eqref{eq:generator-zero-moments} into
\eqref{eq:generator-moment-recurrence},
we obtain, for $n=2k-1$, the two residuals
\begin{equation}\label{eq:kminus1-residuals}
\begin{aligned}
R_0(D)&=-\frac{\delta}{\eta}-\sigma_1D,\\
R_1(D)&=\frac{2\eta D-1}{\eta^2}
-\frac{\delta\sigma_1}{\eta}-\sigma_2D.
\end{aligned}
\end{equation}
For $n=2k$, only $R_1(D)$ occurs.
Thus the prescribed moments satisfy two recurrence
conditions when $r=k-1$, one when $r=k$,
and none when $r\ge k+1$.
The existence of nonzero representing weights
will be examined in the following subsections.

\subsection{\texorpdfstring{The case $r=k-1$}{The case r=k-1}}

Here $n=2k-1$, so the moments in
\eqref{eq:generator-zero-moments}
satisfy the two recurrence relations precisely when
$R_0(D)=R_1(D)=0$.

To obtain a uniform lower bound on the attainable
hull dimension, consider the nonzero weights
$\boldsymbol\lambda=x\boldsymbol u$, where
$\boldsymbol u=(u_1,\ldots,u_n)$ and
$x\in\mathbb F_q^*$.
By Lemma~\ref{lem:lagrange-moments},
\begin{equation}\label{eq:kminus1-terminal-block}
K_D(x\boldsymbol u)
=\operatorname{diag}\!\left(
\mathbf0_{(k-2)\times(k-2)},
\begin{pmatrix}
1+\eta(\eta\Lambda_2+2)x&\delta-\eta\sigma_1x\\
\delta-\eta\sigma_1x&x+D
\end{pmatrix}\right).
\end{equation}
Thus $h^{(k-2)}(\tau)\ge k-2$ for every fixed $\tau$.
The determinant of the lower-right block is
$\Psi_D(x)$, where
\begin{equation}\label{eq:kminus1-psi}
\begin{aligned}
\Psi_D(X)
&=[1+\eta(\eta\Lambda_2+2)X](X+D)
 -(\delta-\eta\sigma_1X)^2\\
&=\eta(2-\eta\sigma_2)X^2
 +[1+\eta(\eta\Lambda_2+2)D
 +2\delta\eta\sigma_1]X+\tau.
\end{aligned}
\end{equation}
Hence $\Psi_D(x)=0$ implies that
$K_D(x\boldsymbol u)$ has rank at most one.
For these weights, Lemma~\ref{lem:lagrange-moments}
also gives
$\nu_0=x\sum_{i=1}^n u_i=0$.

When $(R_0(D),R_1(D))\ne(0,0)$,
rank zero is impossible.
For $k\ge5$, we therefore classify rank-one
generator Gram matrices according to whether
$\nu_0=0$ or $\nu_0\ne0$, leading to two distinct
attainability criteria.

\begin{theorem}\label{thm:kminus1-exact-maxima}
Assume $r=k-1$ and $k\ge4$, and fix
$\tau\in\mathbb F_q^*$.
Then $h^{(k-2)}(\tau)\ge k-2$.
Moreover, $h^{(k-2)}(\tau)=k$ if and only if
$R_0(D)=R_1(D)=0$,
or, equivalently,
\begin{equation}\label{eq:last-rank-one-conditions}
D\ne0,\qquad
\delta+\eta\sigma_1D=0,\qquad
D\eta(\eta\Lambda_2+2)=1.
\end{equation}
Under these conditions, the normalized generator and dual weights
giving hull dimension $k$ are uniquely determined by
\[
\lambda_i=-Du_i,\qquad
\rho_i=-\frac{\tau}{D}u_i,
\qquad 1\le i\le n.
\]

If $k\ge5$ and $(R_0(D),R_1(D))\ne(0,0)$, then
$h^{(k-2)}(\tau)=k-1$ if and only if at least one of the
following conditions holds.
\begin{enumerate}[label=\textup{(\roman*)},leftmargin=*]
\item $R_0(D)\ne0$ and $Q(\beta)\ne0$, where
$\beta=R_1(D)/R_0(D)$. If $D\ne0$, also require
\[
Q\left(\beta+\frac{R_0(D)}{D}\right)\ne0.
\]
\item There exists $x\in\mathbb F_q^*$ such that $\Psi_D(x)=0$.
\end{enumerate}
If $k\ge5$, $(R_0(D),R_1(D))\ne(0,0)$, and neither
condition~\textup{(i)} nor condition~\textup{(ii)} holds, then
$h^{(k-2)}(\tau)=k-2$.
\end{theorem}

\begin{proof}
The weights in~\eqref{eq:kminus1-terminal-block} are nonzero and give
rank at most two, proving $h^{(k-2)}(\tau)\ge k-2$.
By Lemma~\ref{lem:generator-zero-target},
rank zero requires the moment values in
\eqref{eq:generator-zero-moments}.
The first $n$ moment equations and
Lemma~\ref{lem:lagrange-moments}
uniquely force $\lambda_i=-Du_i$.

The two remaining equations hold if and only if $R_0(D)=R_1(D)=0$.
Substituting $\delta=-\eta\sigma_1D$ from the first equation into
the second gives $D\eta(\eta\Lambda_2+2)=1$.
This also implies $D\ne0$, so the weights are nonzero and
\eqref{eq:last-rank-one-conditions} follows.
The dual weights follow from~\eqref{eq:kminus1-gram-correspondence}.

Now assume $k\ge5$ and $(R_0(D),R_1(D))\ne(0,0)$, so rank zero
is excluded. We classify rank-one matrices $K_D(\boldsymbol\lambda)$
with $\boldsymbol\lambda\in(\mathbb F_q^*)^n$.
Index the rows of $K_D$ by $0,\ldots,k-1$, in the order of the
basis in~\eqref{eq:generator-lower}.

First suppose that $K_D$ has rank one with $\nu_0\ne0$, and put
$c=\nu_0$.
Write $K_D=c\boldsymbol z\boldsymbol z^{\mathsf T}$ with $z_0=1$,
and put $\beta=z_1$.
The identities $(K_D)_{1,j}=(K_D)_{0,j+1}$ for
$0\le j\le k-4$ give $z_j=\beta^j$ for $0\le j\le k-3$,
and hence $\nu_s=c\beta^s$ for $0\le s\le2k-6$.
Since $k-1\le2k-6$, the entry
$(K_D)_{0,k-1}=\nu_{k-1}$ gives $z_{k-1}=\beta^{k-1}$.
The entries in the last column and rows $0,\ldots,k-3$ extend
these moment identities through degree $2k-4$.
The entries $(K_D)_{0,k-2}=\nu_{k-2}+\eta\nu_k$ and
$(K_D)_{k-3,k-2}=\nu_{2k-5}+\eta\nu_{2k-3}$ then give
$z_{k-2}=\beta^{k-2}+\eta\beta^k$ and
$\nu_{2k-3}=c\beta^{2k-3}$.
It follows that $\boldsymbol z=g(\beta)$. By the definition of
$K_D[\cdot]$, subtracting $c\beta^j$ from entry $j$ of the moment
sequence subtracts $c\,g(\beta)g(\beta)^{\mathsf T}$ from the
matrix. The resulting matrix is zero. Lemma~\ref{lem:generator-zero-target}
therefore forces
\[
\nu_j=t_j+c\beta^j\qquad(0\le j\le2k),
\]
where $\boldsymbol t$ is the target in~\eqref{eq:generator-zero-moments}.
Conversely, any sequence of this form gives the formal matrix
$c\,g(\beta)g(\beta)^{\mathsf T}$, of rank one because $c\ne0$
and the first entry of $g(\beta)$ is one.

It remains to decide whether this sequence has nonzero representing
weights. Substitution of $t_j+c\beta^j$ in the two recurrences
adds $cQ(\beta)$ and $c\beta Q(\beta)$ to the residuals of
$\boldsymbol t$, respectively. By
Proposition~\ref{prop:realizable-moment-recurrence}, a representation
on $\mathcal A$, allowing zero weights for now, exists precisely when
\[
R_0(D)+cQ(\beta)=0,\qquad R_1(D)+c\beta Q(\beta)=0.
\]
Since the residual pair is nonzero, these equations hold if and only if
\[
R_0(D)\ne0,\qquad Q(\beta)\ne0,\qquad
\beta=\frac{R_1(D)}{R_0(D)},\qquad
c=-\frac{R_0(D)}{Q(\beta)}.
\]
For the first $n$ entries, the target contributes weights $-Du_i$
and the geometric sequence contributes $c\,u_iQ(\beta)/(\beta-\alpha_i)$
by~\eqref{eq:external-evaluation-moments}. Their sum gives the unique
representing weights
\begin{equation}\label{eq:kminus1-finite-weights}
\lambda_i=-u_i\left(D+\frac{R_0(D)}{\beta-\alpha_i}\right).
\end{equation}
For $D=0$, these weights are all nonzero because $R_0(D)\ne0$
and $\beta\notin\mathcal A$.
For $D\ne0$, a weight vanishes exactly when
$\alpha_i=\beta+R_0(D)/D$, so nonvanishing is equivalent to
$Q(\beta+R_0(D)/D)\ne0$.
This gives condition~\textup{(i)}.

Suppose instead that $K_D$ has rank one with $\nu_0=0$.
The rank-one factorization used in the proof of
Theorem~\ref{thm:hankel-distance-three-row} shows that a zero
diagonal entry forces its row to vanish. Thus row $0$ vanishes.
For $1\le j\le k-4$, both
$(K_D)_{j,j}$ and $(K_D)_{j-1,j+1}$ equal $\nu_{2j}$.
If row $j-1$ is zero, so is the diagonal entry in row $j$,
and hence row $j$ itself. This proves that rows $0,\ldots,k-4$
vanish. To reach the next row, use $k\ge5$ and
\[
(K_D)_{k-3,k-3}=\nu_{2k-6}=(K_D)_{k-5,k-1}=0,
\]
so row $k-3$ also vanishes. The entries in these rows give
$\nu_0=\cdots=\nu_{2k-4}=0$, and the entry $(k-3,k-2)$ gives
$\nu_{2k-3}=0$.
Writing $x=\nu_{2k-2}$, these first $n$ moments have unique
weights $\lambda_i=xu_i$ by Lemma~\ref{lem:lagrange-moments}.
Since every $\lambda_i$ is nonzero, $x\ne0$.
By~\eqref{eq:kminus1-terminal-block}, and since rank zero is
excluded, this case occurs if and only if $\Psi_D(x)=0$,
which is condition~\textup{(ii)}.

If neither condition holds, the minimum rank of the generator
Gram matrix is at least two.
The weights in~\eqref{eq:kminus1-terminal-block} attain this bound,
so~\eqref{eq:generator-hull-maximum} gives
$h^{(k-2)}(\tau)=k-2$.
\end{proof}

Although the rank-one classification for fixed $\tau$
in Theorem~\ref{thm:kminus1-exact-maxima} requires
$k\ge5$, the maximum over $\tau$ can be determined
for $k\ge4$ using only the criterion for
$K_D(\boldsymbol\lambda)=0$ and the family
$K_D(x\boldsymbol u)$ in
\eqref{eq:kminus1-terminal-block}.

\begin{corollary}\label{cor:max-hull-kminus1}
Let $k\ge4$ and $r=k-1$. Then
\[
h_{\max}^{(k-2)}=
\begin{cases}
k,&
\eta\Lambda_2+2\ne0,\
\delta(\eta\Lambda_2+2)+\sigma_1=0,\
2-\eta\sigma_2\ne0,\\
k-1,&\text{otherwise}.
\end{cases}
\]

If $h_{\max}^{(k-2)}=k$, then
$h^{(k-2)}(\tau)=k$ only for
\[
\tau^*=
\frac{2-\eta\sigma_2}
{\eta(\eta\Lambda_2+2)^2}.
\]
\end{corollary}

\begin{proof}
By Theorem~\ref{thm:kminus1-exact-maxima},
hull dimension $k$ is equivalent to
$D\eta(\eta\Lambda_2+2)=1$ and
$\delta+\eta\sigma_1D=0$.
Solving gives
\[
\begin{gathered}
\eta\Lambda_2+2\ne0,\qquad
D=\frac{1}{\eta(\eta\Lambda_2+2)},\\
\delta=-\frac{\sigma_1}{\eta\Lambda_2+2},\\
\tau=D-\delta^2
=\frac{\eta\Lambda_2+2-\eta\sigma_1^2}
{\eta(\eta\Lambda_2+2)^2}
=\frac{2-\eta\sigma_2}
{\eta(\eta\Lambda_2+2)^2}.
\end{gathered}
\]
Thus $\tau\ne0$ is equivalent to
$2-\eta\sigma_2\ne0$, proving
the first case and the formula for $\tau^*$.

To obtain hull dimension at least $k-1$,
it suffices to find $x,\tau\in\mathbb F_q^*$
with $\Psi_{\delta^2+\tau}(x)=0$.
Put
\[
p(X)=\eta(2-\eta\sigma_2)X
+1+\eta(\eta\Lambda_2+2)\delta^2
+2\delta\eta\sigma_1.
\]
For $D=\delta^2+\tau$,
equation~\eqref{eq:kminus1-psi} becomes
\[
\Psi_D(X)
=Xp(X)+\tau[1+\eta(\eta\Lambda_2+2)X].
\]
If $p$ is not the zero polynomial,
choose $x\in\mathbb F_q^*$ such that
\[
p(x)[1+\eta(\eta\Lambda_2+2)x]\ne0.
\]
Each factor is a nonzero polynomial of degree
at most one, so at most two nonzero field
elements are excluded.
Such an $x$ exists because $q\ge n=2k-1\ge7$.
Then
\[
\tau=
-\frac{xp(x)}
{1+\eta(\eta\Lambda_2+2)x}
\in\mathbb F_q^*
\]
gives $\Psi_D(x)=0$.

If $p$ is the zero polynomial, then its
linear coefficient gives $2-\eta\sigma_2=0$.
If also $\eta\Lambda_2+2=0$, the identity
\[
\eta\Lambda_2+2
=\eta\sigma_1^2+(2-\eta\sigma_2)
\]
would force $\sigma_1=0$, making the
constant coefficient of $p$ equal to one,
a contradiction.
Hence $\eta\Lambda_2+2\ne0$.
Taking
\[
x=-\frac{1}{\eta(\eta\Lambda_2+2)}
\]
gives $1+\eta(\eta\Lambda_2+2)x=0$
and $\Psi_D(x)=0$ for every
$\tau\in\mathbb F_q^*$.

In either case, a nonzero $\tau$ gives
a generator Gram matrix of rank at most one.
When the criterion for hull dimension $k$
fails, the maximum is therefore $k-1$.
\end{proof}

\subsection{\texorpdfstring{The range $r\ge k$}
{The range r >= k}}

For $r\ge k$, we examine whether the moments
in~\eqref{eq:generator-zero-moments}
can be represented by nonzero weights.
When $r=k$, the first $n=2k$ moments uniquely
determine the weights, while the remaining moment
must satisfy one recurrence relation.
When $r=k+1$, all $n=2k+1$ moments uniquely
determine the weights.
For $r\ge k+2$, the linear map
$\boldsymbol\lambda\mapsto(\nu_0,\ldots,\nu_{2k})$
is surjective with a nontrivial kernel.

Apply the interpolation formula in
Proposition~\ref{prop:realizable-moment-recurrence} to the first $n$
entries of the target~\eqref{eq:generator-zero-moments}. For $n=2k$
its last two entries are $-D$ and $-\delta/\eta$, giving the linear
polynomial $U_D$. For $n=2k+1$ there is one further entry, and the
same formula gives the quadratic expression $V_D$. With $R_0,R_1$
from~\eqref{eq:kminus1-residuals}, these polynomials are
\begin{equation}\label{eq:generator-polynomials}
U_D(z)=R_0(D)-Dz,\qquad
V_D(z)=R_1(D)+R_0(D)z-Dz^2.
\end{equation}
For $r=k$, the weights representing the first
$n=2k$ moments are uniquely given by
$\lambda_i=u_iU_D(\alpha_i)$, and the remaining
recurrence has residual $R_1(D)$.
For $r=k+1$, the weights representing all
$n=2k+1$ moments are uniquely given by
$\lambda_i=u_iV_D(\alpha_i)$.
Since $u_i\ne0$, the nonzero-weight conditions
are determined by the values of $U_D$ and $V_D$
at the evaluation points.

\begin{theorem}\label{thm:generator-maxima}
Let $k\ge4$ and $r=n-k\ge k$, and fix $\tau\in\mathbb F_q^*$ with
$D=\delta^2+\tau$.
\begin{enumerate}[label=\textup{(\roman*)},beginpenalty=10000]
\item Suppose $r=k$. Then $h^{(k-2)}(\tau)=k$ if and only if
\[
R_1(D)=0\quad\text{and}\quad
U_D(\alpha_i)\ne0\ \text{for every }i.
\]
If $R_1(D)=0$ and $U_D(\alpha_j)=0$ for some $j$, then
$h^{(k-2)}(\tau)=k-1$.
If $R_1(D)\ne0$, then $h^{(k-2)}(\tau)\le k-1$, with equality
whenever there exists $\beta\in\mathbb F_q\setminus\mathcal A$ such that
\[
U_D(\alpha_i)(\beta-\alpha_i)\ne R_1(D)
\qquad\text{for every }1\le i\le n.
\]
\item If $r=k+1$, then
\[
h^{(k-2)}(\tau)=
\begin{cases}
k,&V_D(\alpha_i)\ne0\ \text{for every }i,\\
k-1,&\text{otherwise}.
\end{cases}
\]
\item If $r\ge k+2$ and $q>n$, then
$h^{(k-2)}(\tau)=k$.
\end{enumerate}
In particular, for $r=k$ and $q>2n$, one has
$h^{(k-2)}(\tau)=k-1$ whenever the criterion for
$h^{(k-2)}(\tau)=k$ fails.
\end{theorem}

\begin{proof}
Suppose first that $r=k$. By the Lagrange identities, the weights
$\lambda_i=u_iU_D(\alpha_i)$
represent the first $n=2k$ prescribed moments.
They are unique because the Vandermonde system is nonsingular.
The remaining recurrence has residual $R_1(D)$.
Together with Lemma~\ref{lem:generator-zero-target}, this proves
the criterion for rank zero with nonzero weights.

If $R_1(D)=0$ and $U_D(\alpha_j)=0$, then $D\ne0$.
Otherwise $\tau\ne0$ gives $\delta\ne0$ and
$U_D=-\delta/\eta\ne0$.
Consequently, $U_D$ has exactly one zero among the evaluation points.
Replacing this zero weight by any nonzero value makes all weights
nonzero and adds a matrix of rank one to the zero
Gram matrix. Hence $h^{(k-2)}(\tau)=k-1$.

If $R_1(D)\ne0$, Lemma~\ref{lem:moment-perturbation}(ii), with
$R=R_1(D)$, gives the nonzero weights
\[
\lambda_i=u_i\left(U_D(\alpha_i)
-\frac{R_1(D)}{\beta-\alpha_i}\right)
\]
under the stated condition on $\beta$.
For $0\le s\le2k$, the resulting moments equal
the values prescribed in
\eqref{eq:generator-zero-moments} plus $c\beta^s$,
where $c=-R_1(D)/Q(\beta)\ne0$.
The perturbation formula in Section~\ref{sec:hankel} gives a Gram
matrix of rank at most one. Its upper-left entry is $c\ne0$, so
its rank is one.
Lemma~\ref{lem:moment-perturbation}(ii) also guarantees the required
$\beta$ whenever $q>2n$.

Next suppose that $r=k+1$. The unique representation of the
prescribed moments has weights
$u_iV_D(\alpha_i)$, so hull dimension $k$ is attained exactly
when they are all nonzero. Suppose this condition fails.
If $D\ne0$, take $x=-\tau/(\eta^2D)\in\mathbb F_q^*$.
Since $n=2k+1$, Lemma~\ref{lem:lagrange-moments} gives
\[
K_D(x\boldsymbol u)
=\operatorname{diag}\!\left(
\mathbf0_{(k-2)\times(k-2)},
\begin{pmatrix}
\delta^2/D&\delta\\
\delta&D
\end{pmatrix}\right),
\]
which has rank one.
If $D=0$, then $\tau=-\delta^2\ne0$, so
$V_D(z)=R_1(D)-(\delta/\eta)z$ is linear with nonzero coefficient.
It therefore has exactly one zero among the evaluation points.
Replacing the corresponding zero weight by a nonzero value gives
rank one, as in part~\textup{(i)}.
Thus $h^{(k-2)}(\tau)=k-1$ in either case.

Finally, if $r\ge k+2$, then $2k\le n-2$.
Lemma~\ref{lem:weighted-moment}, with $d=2k$
and $q>n$, gives nonzero weights representing the prescribed moments.
Thus the minimum rank of the generator Gram matrix is zero.
\end{proof}

\section{A non-GRS MDS family and monomial inequivalence}\label{sec:codes}

We apply Theorem~\ref{thm:first-constrained-lower} at length
$2k$ to construct non-GRS MDS codes with maximal hull dimension
$k-1$.
We then compute the maximal hull dimension of twice-extended TGRS
codes and the Schur squares of TRL codes with $\ell=k-2$
and $\ell=k-1$.
The maximal hull dimensions establish monomial inequivalence even
for MDS codes with identical code parameters and Schur-square dimensions.

\subsection{\texorpdfstring{A family with maximal hull dimension $k-1$}{A family with maximal hull dimension k-1}}
\label{sec:mds-family}

For the following family, we determine
$h^{(k-2)}(\tau)$ for every $\tau\in\mathbb F_q^*$
and give explicit coordinate multipliers attaining
$h_{\max}^{(k-2)}$.

\begin{theorem}\label{thm:explicit-nongrs-mds-family}
\label{thm:explicit-constrained-nonvacuity}

Let $q_0\ge16$ be a power of two, and let
$q=q_0^m$ for an integer $m\ge3$.
Set
\[
k=\frac{q_0}{2},\qquad n=q_0-2=2k-2.
\]

Choose $c\in\mathbb F_{q_0}^*$, and let
$\boldsymbol\alpha=(\alpha_1,\ldots,\alpha_n)$
be an ordering of the elements of
$\mathbb F_{q_0}\setminus\{0,c\}$.
Choose $\eta\in\mathbb F_q$ such that
$1,\eta^2,\eta^3$ are linearly independent
over $\mathbb F_{q_0}$.
Put $\delta=\eta^2$ and define
\[
\mathcal C=
\operatorname{TRL}_{k,n+2}
(\boldsymbol\alpha,\eta,\eta^2,k-2).
\]

Then $1+c\eta^3\ne0$, and
\[
D^*=\frac{\eta^4}{1+c\eta^3},
\qquad
\tau^*=\frac{c\eta^7}{1+c\eta^3}
\]
belong to $\mathbb F_q^*$. The following statements hold.
\begin{enumerate}[label=\textup{(\roman*)}]
\item The code $\mathcal C$ is a non-GRS $[2k,k,k+1]_q$ MDS code.

\item For both the Euclidean hull over $\mathbb F_q$
and the Hermitian hull after scalar extension to
$\mathbb F_{q^2}$, the fixed-$\tau$ maximum satisfies
\begin{equation}\label{eq:family-fixed-tau}
h^{(k-2)}(\tau)=
\begin{cases}
k-1,&\tau=\tau^*,\\
k-2,&\tau\ne\tau^*,
\end{cases}
\end{equation}
for every $\tau\in\mathbb F_q^*$.
Thus $h_{\max}^{(k-2)}=k-1$.

\item With $L_i(D)=\eta^2+\eta(c-\alpha_i)D$, one has
$L_i(D^*)\ne0$ for every $1\le i\le n$.
Define
\begin{equation}\label{eq:family-multipliers}
a^*=(\tau^*)^{q/2},
\qquad
v_i^*=\left(\frac{\eta(D^*)^2u_i}{L_i(D^*)}\right)^{q/2},
\qquad 1\le i\le n.
\end{equation}
Then
$\mathcal C\operatorname{diag}(v_1^*,\ldots,v_n^*,a^*,1)$
has Euclidean hull dimension $k-1$, and its scalar extension to
$\mathbb F_{q^2}$ has Hermitian hull dimension $k-1$.
\end{enumerate}
\end{theorem}

\begin{proof}
We first verify the MDS and non-GRS properties in (i).
For a subset $I$ of evaluation indices, write
$\boldsymbol\alpha_I=(\alpha_i)_{i\in I}$ and
$Q_I(x)=\prod_{i\in I}(x-\alpha_i)$. Consider a $k\times k$ minor
of~\eqref{eq:generator-lower} with $v_i=1$. With no appended column,
reduce $x^k$ modulo $Q_I$, where $|I|=k$. The coefficient of
$x^{k-2}$ in the remainder is $-e_2(\boldsymbol\alpha_I)$.
With one appended column, expand along that column and let $J$
be the $k-1$ evaluation indices. The coefficients of $x^{k-2}$ in
$x^{k-1}\bmod Q_J$ and $x^k\bmod Q_J$ are
$e_1(\boldsymbol\alpha_J)$ and
$e_1(\boldsymbol\alpha_J)^2-e_2(\boldsymbol\alpha_J)$.
After removing the nonzero Vandermonde factors, the three
conditions for these minors to be nonzero are~\cite{lei2026trl}
\[
\begin{gathered}
1-\eta e_2(\boldsymbol\alpha_I)\ne0,\\
1+\eta\bigl(e_1(\boldsymbol\alpha_J)^2
 -e_2(\boldsymbol\alpha_J)\bigr)\ne0,\\
\eta^2\bigl[1+\eta\bigl(e_1(\boldsymbol\alpha_J)^2
 -e_2(\boldsymbol\alpha_J)\bigr)\bigr]
 -e_1(\boldsymbol\alpha_J)\ne0.
\end{gathered}
\]
Minors using both appended columns are nonzero Vandermonde minors.
All subset symmetric sums lie in $\mathbb F_{q_0}$.
The independence assumption implies $\eta\notin\mathbb F_{q_0}$,
so the first two expressions cannot vanish.
The third expression is a linear combination of
$1,\eta^2,\eta^3$ over $\mathbb F_{q_0}$ with coefficient one
on $\eta^2$, so it is also nonzero.
Thus $\mathcal C$ is MDS.

Let $\beta_1,\ldots,\beta_{2k}$ be the evaluation points of a GRS
code and let $\widetilde u_i$ be their Lagrange weights. In even characteristic one may choose column
multipliers $\sqrt{\widetilde u_i}$. The Gram entries of the resulting
GRS generator are $\sum_i\widetilde u_i\beta_i^{a+b}=0$ for
$0\le a,b<k$, by Lemma~\ref{lem:lagrange-moments} applied to these
$2k$ points. The resulting dimension-$k$ code is
self-dual. Thus every $[2k,k]_q$ GRS code has maximal Euclidean hull
dimension $k$~\cite{chen2023variation}.
For $\mathcal C$, maximizing the bound
\eqref{eq:balanced-hull-bound} over $\tau$ gives
$h_{\max}^{(k-2)}\le k-1$.
The monomial invariance of this maximum proves that
$\mathcal C$ is non-GRS and completes part~\textup{(i)}.

To establish (ii), use $\mathcal A=\mathbb F_{q_0}\setminus\{0,c\}$
to obtain
\[
Q(x)=\frac{x^{q_0-1}-1}{x-c}
=x^{q_0-2}+cx^{q_0-3}+c^2x^{q_0-4}+\cdots.
\]
Thus $\sigma_1=c$, $\sigma_2=c^2$, and $\Lambda_2=0$.
Since the characteristic is two and $\delta=\eta^2$,
\[
\Phi(D)=(1+c\eta^3)D+\eta^4.
\]
The independence of $1,\eta^2,\eta^3$ gives $1+c\eta^3\ne0$.
Consequently, $\Phi(D)$ has the unique root $D^*\ne0$, which
corresponds to
\[
\tau^*=D^*+\eta^4
=\frac{c\eta^7}{1+c\eta^3}\ne0.
\]
For every $1\le i\le n$,
\[
L_i(D^*)=
\frac{\eta^2(1+\alpha_i\eta^3)}{1+c\eta^3}\ne0,
\]
because $\alpha_i\in\mathbb F_{q_0}^*$ and $\eta^3\notin\mathbb F_{q_0}$.
Theorem~\ref{thm:first-constrained-lower}(i) now gives
$h^{(k-2)}(\tau^*)=k-1$ and excludes this value for
$\tau\ne\tau^*$. Since $q=q_0^m>2n$,
Theorem~\ref{thm:first-constrained-lower}\textup{(ii)} gives
$h^{(k-2)}(\tau)=k-2$ for every $\tau\ne\tau^*$,
proving \textup{(ii)} in both hull settings.

Finally, we verify the explicit multipliers in (iii).
Since $x\mapsto x^{q/2}$ inverts squaring on $\mathbb F_q$,
the multipliers in~\eqref{eq:family-multipliers} satisfy
$(a^*)^2=\tau^*$ and have normalized dual weights
\[
\rho_i^*:=\frac{\tau^*u_i^2}{(v_i^*)^2}
=\frac{\tau^*u_iL_i(D^*)}{\eta(D^*)^2}
=u_iP_{D^*}(\alpha_i).
\]
Differentiating $x(x-c)Q(x)=x^{q_0}-x$ at $\alpha_i$ gives
$u_i=\alpha_i(\alpha_i-c)$, and direct substitution in
\eqref{eq:linear-target-polynomial} gives
\[
P_{D^*}(z)=c(1+\eta^3z).
\]
Consequently,
\begin{equation}\label{eq:family-explicit-weights}
\rho_i^*=c\alpha_i(\alpha_i-c)(1+\alpha_i\eta^3),
\qquad
(v_i^*)^2=
\frac{\tau^*\alpha_i(\alpha_i-c)}{c(1+\alpha_i\eta^3)}.
\end{equation}
All factors in these expressions are nonzero.
For $\mu_s^*=\sum_i\rho_i^*\alpha_i^s$, Lemma~\ref{lem:lagrange-moments}
now yields
\[
\begin{gathered}
\mu_0^*=\cdots=\mu_{n-3}^*=0,\qquad
\mu_{n-2}^*=c\eta^3,\\
\mu_{n-1}^*=c+c^2\eta^3,\qquad \mu_n^*=c^2.
\end{gathered}
\]
Inserting these moments into the normalized dual Gram matrix gives
the explicit factorization
\begin{equation}\label{eq:family-dual-gram-factorization}
H_k(\boldsymbol\mu^*)+B^{(k-2)}(\tau^*)
=\frac{1}{1+c\eta^3}\,
\boldsymbol z_\eta\boldsymbol z_\eta^{\mathsf T},
\qquad
\boldsymbol z_\eta=
\begin{pmatrix}\boldsymbol0_{k-3}\\ \eta^3\\ 1\\ \eta^2\end{pmatrix}.
\end{equation}
This matrix has rank one. Since the dual dimension is $r+2=k$,
the scaled code in part~\textup{(iii)} has Euclidean hull dimension
$k-1$.
Since the scaled code is defined over $\mathbb F_q$, its
generator matrix $G$ satisfies $GG^\dagger=GG^{\mathsf T}$
after scalar extension to $\mathbb F_{q^2}$.
Scalar extension preserves rank, so the Hermitian hull dimension
is also $k-1$, proving~\textup{(iii)}.
\end{proof}

\begin{remark}
The linear independence condition on
$1,\eta^2,\eta^3$ in Theorem~\ref{thm:explicit-nongrs-mds-family}
can be satisfied for every $m\ge3$.
For $m\ge4$, any element $\eta$ of degree $m$
over $\mathbb F_{q_0}$ satisfies this condition.
For $m=3$, the map $t\mapsto t^3+t$ is not
surjective on $\mathbb F_{q_0}$, since it takes
the same value at $0$ and $1$.
 Choose $b\in\mathbb F_{q_0}$ outside its image. The cubic
$X^3+X+b$ has no root in $\mathbb F_{q_0}$ and is therefore
irreducible. A root $\eta\in\mathbb F_{q_0^3}$ has degree three
and satisfies $\eta^3=\eta+b$. Replacing $\eta$ by $\eta^3$ in the
basis $\{1,\eta^2,\eta\}$ preserves linear independence.
\end{remark}

\subsection{Comparison with twice-extended TGRS codes}
\label{sec:hook-comparison}

The twice-extended TGRS codes of Cheng and
Zhu~\cite[Definition~2.1]{cheng2026twice} are monomially equivalent
to the $\ell=k-1$ specialization of the TRL family.
In the notation of Cheng and Zhu, dividing the last generator row
by $\lambda_1$ and scaling coordinate $n+1$ by $\lambda_1$ gives
$\eta=\lambda_2/\lambda_1$ and
$\delta=\delta_{\mathrm{CZ}}/\lambda_1$, where
$\lambda_1,\lambda_2\in\mathbb F_q^*$.
Conversely, taking $\lambda_1=1$, $\lambda_2=\eta$, and
$\delta_{\mathrm{CZ}}=\delta$ gives the $\ell=k-1$ TRL specialization
with $\eta\ne0$.
For length $2k$, we determine the maximal hull dimensions of the
$\ell=k-1$ TRL codes.

\begin{proposition}\label{prop:hook-kminus1-realization-threshold}
Let $k\ge5$ and $n=2k-2\le q$, and let
$\alpha_1,\ldots,\alpha_n\in\mathbb F_q$ be pairwise distinct.
Fix $\eta\in\mathbb F_q^*$, $\delta\in\mathbb F_q$,
and $\ell=k-1$. Then
\[
h^{(k-1)}(\tau)\le k-2
\qquad\text{for every }\tau\in\mathbb F_q^*,
\]
with equality whenever $q>n$.
\end{proposition}

\begin{proof}
Put $r=k-2$ and write
$F_{k-1}=[\boldsymbol p\mid-\boldsymbol e_{r+1}]$ as in
Lemma~\ref{lem:normalized-parity-check}, where $\boldsymbol e_{r+1}$ is the
last standard basis vector of $\mathbb F_q^{r+2}$, with coordinates
indexed by $0,\ldots,r+1$. Then
\[
B^{(k-1)}(\tau)=\boldsymbol p\boldsymbol p^{\mathsf T}
+\tau\boldsymbol e_{r+1}\boldsymbol e_{r+1}^{\mathsf T},
\]
whose lower-right $3\times3$ block has $b_{00}=\eta^2\ne0$.
Thus Theorem~\ref{thm:hankel-distance-three-row}(i)
excludes rank zero for the dual Gram matrix. Since $r\ge3$,
Theorem~\ref{thm:hankel-distance-three-row}(i) and~\eqref{eq:rank-one-tail}
show that rank one would force
\[
\mu_0=\cdots=\mu_{n-1}=0,\qquad \mu_n=-\tau.
\]
The degree-$n$ recurrence in
Proposition~\ref{prop:realizable-moment-recurrence} would then give
$-\tau=0$, contradicting $\tau\ne0$.
Thus every represented dual Gram matrix has rank at least two,
and~\eqref{eq:fixed-tau-hull-maximum} gives the upper bound.

If $q>n$, choose $\beta\in\mathbb F_q\setminus\mathcal A$
and set
\[
\rho_i=\frac{\tau u_i}{\beta-\alpha_i},\qquad
1\le i\le n.
\]
All these weights are nonzero. Since $n=2r+2$, the identities
in~\eqref{eq:external-evaluation-moments} give
\[
H_{r+2}(\boldsymbol\mu(\boldsymbol\rho))+B^{(k-1)}(\tau)
=\boldsymbol p\boldsymbol p^{\mathsf T}
+\frac{\tau}{Q(\beta)}\,
\boldsymbol h_{r+2}(\beta)\boldsymbol h_{r+2}(\beta)^{\mathsf T}.
\]
This matrix has rank at most two.
The preceding lower bound makes its rank exactly two, giving
$h^{(k-1)}(\tau)=k-2$.
\end{proof}

Let $\mathcal C_{k-2}=\mathcal C$ denote the MDS code in
Theorem~\ref{thm:explicit-nongrs-mds-family}.
Using the same evaluation points and parameters
$\eta,\delta=\eta^2$, define
\[
\mathcal C_{k-1}
=\operatorname{TRL}_{k,n+2}
(\boldsymbol\alpha,\eta,\eta^2,k-1).
\]
By the MDS criterion for $\ell=k-1$ in
\cite[Theorem~3.10]{lei2026trl},
the code $\mathcal C_{k-1}$ is MDS, since
the required nonvanishing conditions follow
from $\alpha_i\in\mathbb F_{q_0}$,
$\delta=\eta^2$, and the fact that $\eta$
has degree at least three over $\mathbb F_{q_0}$.
Together with
Theorem~\ref{thm:explicit-nongrs-mds-family},
this shows that $\mathcal C_{k-2}$ and
$\mathcal C_{k-1}$ are both
$[2k,k,k+1]_q$ MDS codes.

We now compute the Schur squares of TRL codes with
$\ell=k-2$ and $\ell=k-1$ at length $2k$.

\begin{proposition}\label{prop:balanced-schur-squares}
Let $q$ be a prime power, $k\ge4$, and $n=2k-2\le q$.
For pairwise distinct $\alpha_1,\ldots,\alpha_n\in\mathbb F_q$,
$\eta\in\mathbb F_q^*$, and $\delta\in\mathbb F_q$, put
$\mathcal C_\ell=\operatorname{TRL}_{k,n+2}
(\boldsymbol\alpha,\eta,\delta,\ell)$.
For both $\ell=k-2$ and $\ell=k-1$, one has
\[
\mathcal C_\ell^{\star2}=\mathbb F_q^{2k}.
\]
\end{proposition}

\begin{proof}
Write $\boldsymbol e_j=(\alpha_1^j,\ldots,\alpha_n^j,0,0)$ and
$E=\mathbb F_q^n\times\{(0,0)\}$.
We first show that the Schur square contains $E$ for each value of
$\ell$. We then obtain two additional independent directions from
the appended coordinates.
In both generators~\eqref{eq:generator-lower} and
\eqref{eq:generator-upper}, with $v_i=1$, the first $k-2$ rows are
$\boldsymbol e_0,\ldots,\boldsymbol e_{k-3}$.
Their Schur products give $\boldsymbol e_j$ for $0\le j\le2k-6$.

For $\ell=k-2$, products of these rows with the last row give
$\boldsymbol e_j$ for $k-1\le j\le2k-4$.
Since $k\ge4$, these two ranges together cover $0\le j\le2k-4$.
The product of $\boldsymbol e_{k-3}$ with the penultimate row is
$\boldsymbol e_{2k-5}+\eta\boldsymbol e_{2k-3}$.
As $\eta\ne0$, this also gives $\boldsymbol e_{2k-3}$.
Thus $\boldsymbol e_0,\ldots,\boldsymbol e_{n-1}$ belong to
$\mathcal C_{k-2}^{\star2}$.

For $\ell=k-1$, products with the penultimate row first give
$\boldsymbol e_j$ for $k-2\le j\le2k-5$.
Together with the products of the first $k-2$ rows, these cover
$0\le j\le2k-5$.
Products of the last row with $\boldsymbol e_{k-4}$ and
$\boldsymbol e_{k-3}$ then give, successively,
\[
\boldsymbol e_{2k-5}+\eta\boldsymbol e_{2k-4},
\qquad
\boldsymbol e_{2k-4}+\eta\boldsymbol e_{2k-3}.
\]
Hence $\boldsymbol e_0,\ldots,\boldsymbol e_{n-1}$ also belong to
$\mathcal C_{k-1}^{\star2}$.

The evaluation points are distinct, so the Vandermonde matrix shows
that these $n$ vectors span $E$.
In either code, the squares of the last two generator rows have
appended coordinate pairs $(0,1)$ and $(1,\delta^2)$, which are
linearly independent. Their images modulo $E$ therefore add two
dimensions. This proves the assertion.
\end{proof}

In particular, the MDS codes $\mathcal C_{k-2}$
and $\mathcal C_{k-1}$ have the same Schur square,
namely $\mathbb F_q^{2k}$.
Since a $[2k,k]_q$ GRS code has Schur-square
dimension $2k-1$, neither code is monomially
equivalent to a GRS code.

By Theorem~\ref{thm:explicit-nongrs-mds-family}
and Proposition~\ref{prop:hook-kminus1-realization-threshold},
the maximal Euclidean hull dimensions of
$\mathcal C_{k-2}$ and $\mathcal C_{k-1}$
are $k-1$ and $k-2$, respectively.
Hence these two MDS codes are monomially inequivalent,
even though their code parameters and Schur squares
coincide.
More generally, the difference in maximal hull dimensions
establishes monomial inequivalence between every
TRL code constructed in Subsection~\ref{sec:mds-family}
and every $[2k,k]_q$ twice-extended TGRS code
with the same $k$ and $q$.
This inequivalence persists after scalar extension to
$\mathbb F_{q^2}$, since the corresponding maximal
Hermitian hull dimensions remain $k-1$ and $k-2$.

\section{Conclusion}\label{sec:conclusion}
This paper studies maximal hull dimensions within fixed monomial equivalence classes of \([n+2,k]_q\) twisted Roth–Lempel codes with \(\ell=k-2\) and \(\eta\ne0\). We consider Euclidean hulls over \(\mathbb F_q\) for even \(q\) and Hermitian hulls after scalar extension to \(\mathbb F_{q^2}\) for arbitrary prime powers \(q\). By expressing hull dimensions in terms of ranks of normalized Gram matrices, we reduce the fixed-\(\tau\) problem to rank minimization under nonzero evaluation-weight constraints. We use dual Gram matrices for \(2\le r\le k-2\) and generator Gram matrices for \(r\ge k-1\). The resulting analysis, based on representations of prescribed moment sequences by nonzero weights and controlled rank-one perturbations, yields the fixed-\(\tau\) maxima in the parameter ranges and under the field-size assumptions listed in Table 1. In particular, for \(r=k-1\) and \(k\ge5\), we obtain necessary and sufficient conditions for the fixed-\(\tau\) maximum to equal \(k\), \(k-1\), or \(k-2\), under the basic requirement \(q\ge n\). We also determine the maximum over \(\tau\) in the critical parameter ranges. Moreover, for every \(k\ge4\) and \(n>k\), every integer from zero to the fixed-\(\tau\) maximum is attainable while the multipliers of the two appended coordinates remain fixed.

As an application, we use the hull-dimension criterion for \(r=k-2\) to construct explicit non-GRS \([2k,k,k+1]_q\) MDS codes in characteristic two with maximal hull dimension \(k-1\). For the same parameters, every code in the twice-extended TGRS family of Cheng and Zhu has maximal hull dimension \(k-2\), proving that the constructed TRL codes are monomially inequivalent to that family. We also compare the TRL codes with \(\ell=k-2\) and \(\ell=k-1\), obtained from the same evaluation points and remaining defining parameters. Both are \([2k,k,k+1]_q\) MDS codes with Schur-square dimension \(2k\), but their maximal Euclidean hull dimensions are \(k-1\) and \(k-2\), respectively. Thus maximal hull dimension distinguishes MDS codes even when their code parameters and Schur-square dimensions coincide. Their scalar extensions to \(\mathbb F_{q^2}\) retain the same distinction in maximal Hermitian hull dimensions, so the monomial inequivalence also persists over \(\mathbb F_{q^2}\).

Several questions remain for further study. These include extending the classification to parameter ranges beyond those covered in Table 1 and weakening the sufficient field-size assumptions. The Euclidean hull problem in odd characteristic, where square constraints on the normalized weights and on \(\tau\) must be taken into account, is another natural direction for future work.

\appendix
\section{\texorpdfstring{The case $(k,r)=(5,2)$}{The case (k,r)=(5,2)}}\label{app:k5-r2}

For $r=2$, the condition $\rank(H_4(\boldsymbol\mu)+B)=1$ need not
force the first four moments to vanish. We first describe all
moment vectors satisfying this condition and then test whether
their uniquely determined weights are nonzero.

\begin{lemma}\label{lem:r2-rank-one-parameterization}
Let $\mathbb K$ be a field, and let $B\in\mathbb K^{4\times4}$
be symmetric, with rows and columns indexed by $0,\ldots,3$.
Suppose that all entries outside its lower-right $3\times3$
principal block $B_0=(b_{ij})_{0\le i,j\le2}$ are zero and that
$(b_{00},b_{01})\ne(0,0)$.
For $\boldsymbol\mu=(\mu_0,\ldots,\mu_6)\in\mathbb K^7$,
one has $\rank(H_4(\boldsymbol\mu)+B)=1$ if and only if one of the
following cases holds.
\begin{enumerate}[label=\textup{(\roman*)},leftmargin=*]
\item $b_{00}\ne0$ and
\[
\boldsymbol\mu=(0,0,0,0,\xi,\upsilon,\zeta),
\]
where $\xi,\upsilon,\zeta$ are given by~\eqref{eq:rank-one-tail}.
\item There exist $\gamma\in\mathbb K^*$ and $\theta\in\mathbb K$
satisfying
\begin{equation}\label{eq:r2-rank-one-constraint}
\gamma\bigl(b_{00}\theta^2-b_{01}\theta+b_{11}-b_{02}\bigr)
=b_{00}^2
\end{equation}
such that the moments are given successively by
\begin{equation}\label{eq:r2-rank-one-family}
\begin{aligned}
&\mu_0=\gamma,\qquad \mu_1=\gamma\theta,\qquad
 \mu_2=\gamma\theta^2-b_{00},\\
&\mu_3=\theta\mu_2-b_{01},\qquad
 \mu_4=\theta\mu_3-b_{02},\\
&\mu_5=\frac{\mu_2\mu_3}{\gamma}-b_{12},\qquad
 \mu_6=\frac{\mu_3^2}{\gamma}-b_{22}.
\end{aligned}
\end{equation}
In this case $\gamma=\mu_0$ and $\theta=\mu_1/\mu_0$.
\end{enumerate}
\end{lemma}

\begin{proof}
Put $G=H_4(\boldsymbol\mu)+B$.
If $\mu_0=0$ and $G$ has rank one, its principal $2\times2$
minors give $\mu_1=\mu_2=\mu_3=0$.
If $b_{00}=0$, then $b_{01}\ne0$ and the principal minor on
indices $1,2$ has determinant $-b_{01}^2\ne0$, a contradiction.
For $b_{00}\ne0$, using the pivot $G_{11}=b_{00}$ gives exactly
the three moments in~\eqref{eq:rank-one-tail}. Conversely, those
moments make the lower-right block have rank one.

Now suppose that $\gamma=\mu_0\ne0$, and put $\theta=\mu_1/\gamma$.
Rank one is equivalent to
$G_{ij}=G_{i0}G_{0j}/\gamma$ for $1\le i\le j\le3$.
The entries with indices $(1,1)$, $(1,2)$, $(1,3)$, $(2,3)$, and
$(3,3)$ give~\eqref{eq:r2-rank-one-family}.
The remaining entry, with indices $(2,2)$, requires
\[
0=\gamma(\theta\mu_3+b_{11}-b_{02})-\mu_2^2
 =\gamma\bigl(b_{00}\theta^2-b_{01}\theta+b_{11}-b_{02}\bigr)
  -b_{00}^2.
\]
This is~\eqref{eq:r2-rank-one-constraint}. Conversely, these formulas
give
\[
H_4(\boldsymbol\mu)+B
=\frac1\gamma
(\gamma,\gamma\theta,\mu_2,\mu_3)^{\mathsf T}
(\gamma,\gamma\theta,\mu_2,\mu_3),
\]
which has rank one.
\end{proof}

Fix $k=5$, $r=2$, and hence $n=7$, with $\ell=k-2$,
$\eta\in\mathbb F_q^*$, $\delta\in\mathbb F_q$, and pairwise distinct
$\alpha_1,\ldots,\alpha_7\in\mathbb F_q$.
For $\tau\in\mathbb F_q^*$, set $D=\delta^2+\tau$ and
$B=B^{(k-2)}(\tau)$, whose lower-right $3\times3$ block $B_0$ is given
in~\eqref{eq:leading-block-entries}.
If $D\ne0$, then $b_{00}=\eta^2D\ne0$. If $D=0$, then
$\delta\ne0$ and $b_{01}=-\eta\delta\ne0$.
Thus the preceding lemma applies for every $\tau$.

It gives the following explicit finite set $\mathcal M_\tau$
of all moment vectors with $\rank(H_4(\boldsymbol\mu)+B)=1$.
If $b_{00}\ne0$, include the vector in part~\textup{(i)} and, for
each $\theta\in\mathbb F_q$ with
$b_{00}\theta^2-b_{01}\theta+b_{11}-b_{02}\ne0$, include the vector
in~\eqref{eq:r2-rank-one-family} with
\[
\gamma=\frac{b_{00}^2}
 {b_{00}\theta^2-b_{01}\theta+b_{11}-b_{02}}.
\]
If $b_{00}=0$, then $b_{01}\ne0$. Take instead all vectors
in~\eqref{eq:r2-rank-one-family} with
\[
\theta=\frac{b_{11}-b_{02}}{b_{01}},
\qquad \gamma\in\mathbb F_q^*.
\]
Distinct parameter choices in~\eqref{eq:r2-rank-one-family} give
distinct vectors, since $\gamma=\mu_0$ and $\theta=\mu_1/\mu_0$.
Thus $\mathcal M_\tau$ is nonempty. It contains at most $q+1$ vectors
when $b_{00}\ne0$ (including the vector with $\mu_0=0$), and exactly
$q-1$ vectors when $b_{00}=0$.

For $\boldsymbol\mu=(\mu_0,\ldots,\mu_6)\in\mathbb F_q^7$, put
$\sigma_0=1$ and define
\begin{equation}\label{eq:r2-interpolation-polynomial}
P_{\boldsymbol\mu}(z)
=\sum_{d=0}^{6}
 \left(\sum_{j=0}^{d}\sigma_j\mu_{d-j}\right)z^{6-d}.
\end{equation}

\begin{theorem}\label{thm:k5-r2-exact}
Assume $(k,r)=(5,2)$ and $q>n=7$, and fix
$\tau\in\mathbb F_q^*$.
Then $h^{(k-2)}(\tau)=3$ if and only if some
$\boldsymbol\mu\in\mathcal M_\tau$ satisfies
\begin{equation}\label{eq:r2-nonzero-weight-test}
P_{\boldsymbol\mu}(\alpha_i)\ne0
\qquad\text{for every }1\le i\le7.
\end{equation}
Otherwise, $h^{(k-2)}(\tau)=2$.
\end{theorem}

\begin{proof}
For every $\boldsymbol\mu\in\mathbb F_q^7$, the nonsingular
Vandermonde system
\[
\sum_{i=1}^7\rho_i\alpha_i^d=\mu_d,\qquad 0\le d\le6,
\]
has a unique solution $\boldsymbol\rho$.
Polynomial division by $z-\alpha_i$ gives
\[
\frac{Q(z)}{z-\alpha_i}
=\sum_{d=0}^{6}
 \left(\sum_{j=0}^{d}\sigma_j\alpha_i^{d-j}\right)z^{6-d}.
\]
Multiplying by $\rho_i$ and summing over $i$ yields
\[
P_{\boldsymbol\mu}(z)
=\sum_{i=1}^7\rho_i\frac{Q(z)}{z-\alpha_i}.
\]
Evaluating at each $\alpha_i$ gives
$P_{\boldsymbol\mu}(\alpha_i)=\rho_iQ'(\alpha_i)=\rho_i/u_i$.
Hence the weights are $\rho_i=u_iP_{\boldsymbol\mu}(\alpha_i)$
and are nonzero exactly when~\eqref{eq:r2-nonzero-weight-test} holds.
The description of $\mathcal M_\tau$ therefore proves the criterion for attaining
rank one. Rank zero is excluded by
Theorem~\ref{thm:hankel-distance-three-row}.

If no candidate passes the nonvanishing test, choose any
$\boldsymbol\mu\in\mathcal M_\tau$.
Its representing weight vector has a zero coordinate.
Since $q>7$, Lemma~\ref{lem:moment-perturbation}(i) gives nonzero
weights representing $\mu_d+c\beta^d$ for $0\le d\le6$.
Their Gram matrix is
\[
H_4(\boldsymbol\mu)+B
+c\,\boldsymbol h_4(\beta)\boldsymbol h_4(\beta)^{\mathsf T}
\]
and has rank at most two.
Ranks zero and one have been excluded for nonzero weights, so the
minimum rank is two.
The dual has dimension four, and~\eqref{eq:fixed-tau-hull-maximum}
therefore gives the asserted hull dimensions $4-1=3$ and $4-2=2$.
\end{proof}

\end{document}